\documentclass[journal]{IEEEtran}
\usepackage{cite}
\usepackage{color,soul}
\usepackage[rightcaption]{sidecap}
\ifCLASSINFOpdf
   
   \usepackage[pdftex]{graphicx}
   \usepackage{multirow}
   \usepackage{tabularew}
   \usepackage[flushleft]{threeparttable}
   \graphicspath{{./pdf/}{./jpeg/}{./eps/}}
  \DeclareGraphicsExtensions{ ,.jpeg,.png}
\else
   \usepackage[dvips]{graphicx}
   \graphicspath{{./eps/}}
   \DeclareGraphicsExtensions{.eps}
\fi
\usepackage{amsmath}
\usepackage{array}
\ifCLASSOPTIONcompsoc
 \usepackage[caption=false,font=normalsize,labelfont=sf,textfont=sf]{subfig}
\else
  \usepackage[caption=false,font=footnotesize]{subfig}
\fi

\usepackage{stfloats}
\usepackage{url}
  \usepackage{color}

\usepackage{url}
\begin{document}

\title{X-SRAM: Enabling In-Memory Boolean Computations in CMOS Static Random Access Memories}

\author{Amogh Agrawal*, Akhilesh Jaiswal*, Chankyu Lee and Kaushik~Roy,~\IEEEmembership{Fellow,~IEEE}% <-this % stops a space

\IEEEauthorblockA{School of Electrical and Computer Engineering,
Purdue University, West Lafayette, IN-47907, USA\\
(* Equal Contributors)\\
Email: \{agrawa64, jaiswal, lee2216, kaushik\}@purdue.edu}

}
\maketitle
\pagenumbering{gobble}

\begin{abstract}

Silicon-based Static Random Access Memories (SRAM) and digital Boolean logic have been the workhorse of the state-of-the-art computing platforms. Despite tremendous strides in scaling the ubiquitous metal-oxide-semiconductor transistor, the underlying \textit{von-Neumann} computing architecture has remained unchanged. The limited throughput and energy-efficiency of the state-of-the-art computing systems, to a large extent, results from the well-known \textit{von-Neumann bottleneck}. The energy and throughput inefficiency of the von-Neumann machines have been accentuated in recent times due to the present emphasis on data-intensive applications like artificial intelligence, machine learning, cryptography \textit{etc}. A possible approach towards mitigating the overhead associated with the von-Neumann bottleneck is to enable \textit{in-memory} Boolean computations. In this manuscript, we present an augmented version of the conventional SRAM bit-cells, called \textit{the X-SRAM}, with the ability to perform in-memory, vector Boolean computations, in addition to the usual memory storage operations. We propose at least six different schemes for enabling in-memory vector computations including NAND, NOR, IMP (implication), XOR logic gates with respect to different bit-cell topologies $-$ the 8T cell and the 8$^+$T Differential cell. In addition, we also present a novel \textit{`read-compute-store'} scheme, wherein the computed Boolean function can be directly stored in the memory without the need of latching the data and carrying out a subsequent write operation. The feasibility of the proposed schemes have been verified using predictive transistor models and detailed Monte-Carlo variation analysis. As an illustration, we also present the efficacy of the proposed in-memory computations by implementing AES (advanced encryption standard) algorithm on a non-standard von-Neumann machine wherein the conventional SRAM is replaced by X-SRAM. Our simulations indicated that up-to 75\% of memory accesses can be saved using the proposed techniques.

\end{abstract}


% \fi
\IEEEpeerreviewmaketitle

\section{Introduction}

\IEEEPARstart{S}{ince} the invention of transistor switches \cite{bardeen}, there has been an ever-increasing demand for speed and energy-efficiency in computing systems. Almost all the state-of-the-art computing platforms are based on the well-known \textit{von-Neumann} architecture which is characterized by decoupled \textit{memory storage} and \textit{computing cores}. Running data-intensive applications on such von-Neumann machines, like artificial intelligence, search engines, neural networks, biological systems, financial analysis \textit{etc.}, are limited by the \textit{von Neumann bottleneck} \cite{vnbottleneck}. This bottleneck results due to frequent and large amounts of data transfer between the physically separate memory units and compute cores. Moreover, frequent to-and-fro data transfers incur large energy overheads in addition to limiting the overall throughput.

\begin{figure}[t]
\centering
\includegraphics[width=0.45\textwidth]{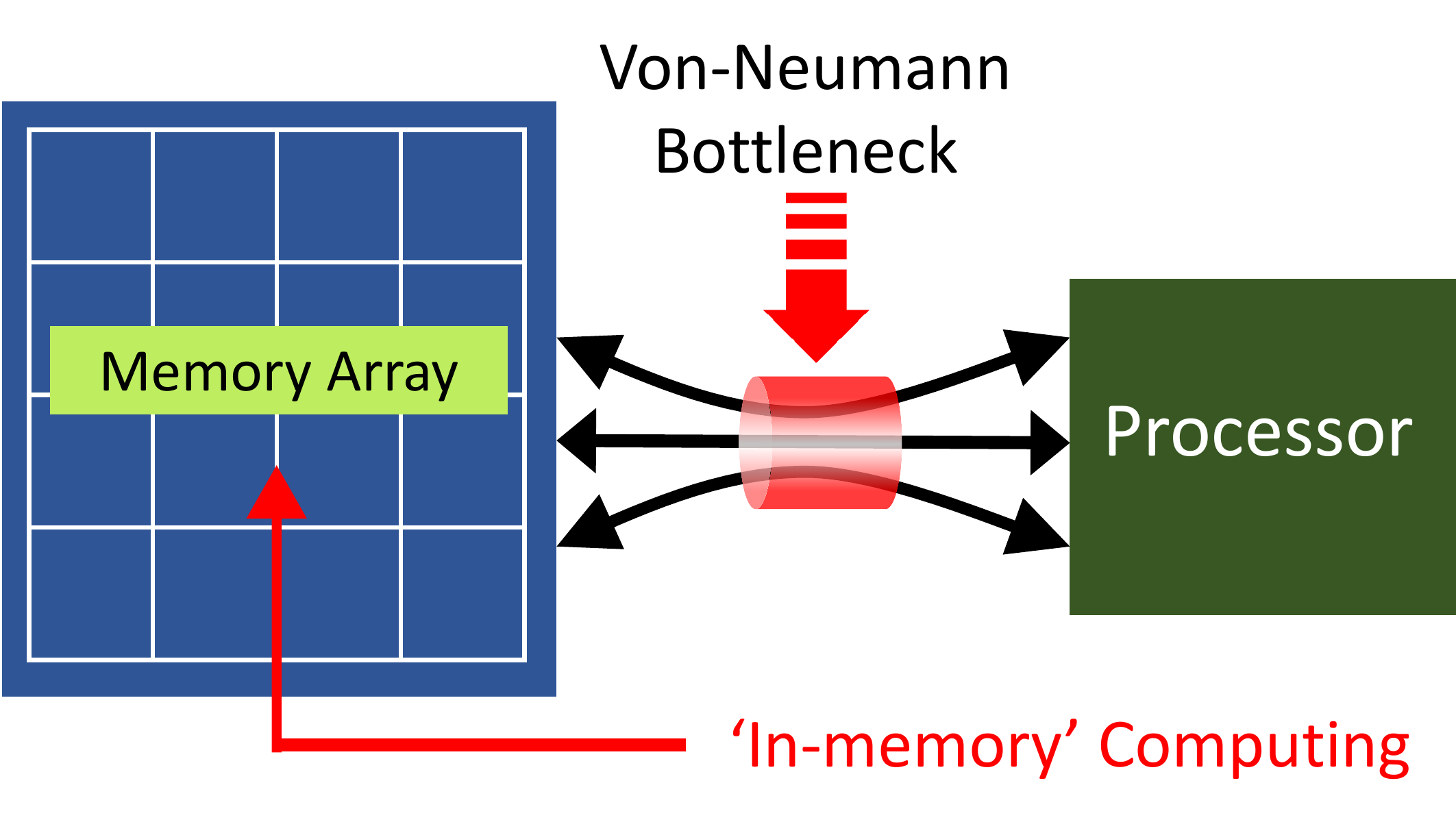}
\caption{Illustration of the von-Neumann bottleneck. Frequent to-and-fro data transfers between the processor and memory units incur large energy consumption and limits the throughput. Computing within the memory array enhances the memory functionality thereby reducing the number of unnecessary transfers of data for certain class of operations like vector bit-wise Boolean logic \textit{etc}.}
\label{bottleneck} 
\end{figure}

In order to overcome the von-Neumann bottleneck, there have been many efforts to develop new computing paradigms. One of the most promising approach is the \textit{in-memory computing}, which aims to embed logic within the memory array in order to reduce memory-processor data transfers. Conceptually, the in-memory compute paradigm is illustrated in Fig. \ref{bottleneck}. It shows two physically separated blocks $-$ the processor and the memory unit and the associated computing bottleneck. In-memory techniques tend to bypass the von-Neumann bottleneck by accomplishing computations right inside the memory array, as shown in the figure. In other words, in-memory-compute blocks store data exactly like a standard memory, however, they enable additional operations without expensive area or energy overheads. By enabling logic computations in-memory, significant improvements, both in energy efficiency and throughput are expected \cite{rsnm6t,cc,shanbhag,Kang_2015}.

% for example, computational Random Access Memories (RAMs) \cite{cram}, content addressable memories (CAM) \cite{cam} and compute memories \cite{shanbhag}, among others. These new paradigms, commonly known as \textit{in-memory computing}, aim to embed logic within the memory array in order to reduce memory-processor data transfers. In-memory-compute blocks store data exactly like a standard memory. However, they not only support regular memory transactions, but enable additional operations without expensive area or energy overheads. By enabling logic computations as close to the memory units as possible, significant improvements, both in energy efficiency and compute throughput are expected [Ref]-[Ref].

%\begin{figure}[t]
%\centering
%\includegraphics[width=0.35\textwidth]{proposal.pdf}
%\caption{A summary of \textit{In-Memory} computing schemes proposed in this work.}
%\label{summary}
%\end{figure}
 
 \begin{figure}[t]
\centering
\includegraphics[width=0.5\textwidth]{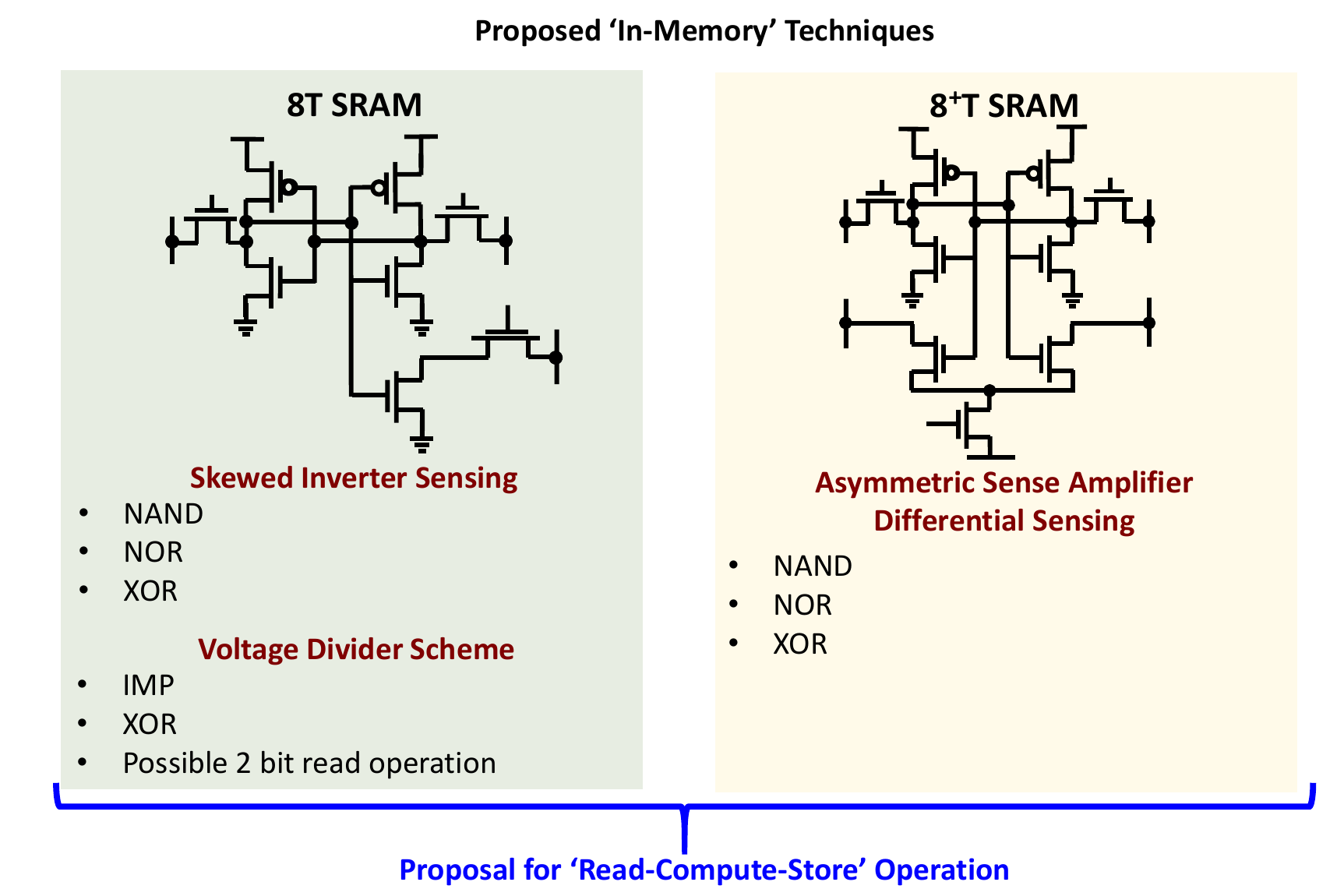}
\caption{A summary of \textit{In-Memory} computing schemes proposed in this work. With respect to the 8T cell, we present bit-wise NAND, NOR and XOR operations using skewed inverter sensing. Further, we present the voltage-divider based operation of 8T-cells for IMP and XOR gates. With respect to the 8$^+$T-cells, we present bit-wise NAND, NOR and XOR operations using asymmetric differential SAs. Moreover, a \textit{`read-compute-store'} operation has been presented for both types of bit-cells.}
\label{summary} 
\end{figure}
 
Due to the potential impact of in-memory computing on future computing platforms, various proposals spanning right from conventional complementary metal-oxide semiconductor (CMOS) to beyond-CMOS technologies can be found in the literature. For example, Ref. \cite{cram} proposed integrating an ALU (arithmetic-logic-unit) close to the memory unit to exploit the wide memory bandwidth, while Ref. \cite{rsnm6t} reconfigures a standard 6 transistor (6T) static random-access memory (SRAM) cells as content addressable memories (CAMs) and enable bit-wise logical operations. 6T-SRAM cells have also been used to implement machine learning classifiers \cite{6tml}, and dot-products in analog domain for pattern recognition \cite{shanbhag}. The underlying idea is to enable multiple rows of memory bit-cells and directly read out a voltage at the pre-charged bit-lines corresponding to the desired operation. However, the 6T-SRAM bit-cells have a coupled read-write path that imposes conflicting constraints on the design of the 6T cell, thereby raising issues of read-disturb failures. Moreover, activating multiple word-lines may cause short-circuit paths, thereby flipping the cell states nondeterministically. The read-disturb failure is further accentuated by the fact that once the BL has discharged, activating subsequent word-lines perform a \textit{pseudo-write} operation on the 6T cell, given the shared read-write path. A 6T-SRAM based on the deeply depleted channel (DDC) technology \cite{6tddc} was recently proposed for searching and in-memory computing applications, which had decoupled read-write paths. However, all of these proposals perform the computation in the peripheral circuits and read out the data. A subsequent memory-write operation is required to store the data back in the memory array. Thus, in our work, we use standard CMOS 8T- and 8$^+$T Differential SRAM cells due to their decoupled read-write mechanisms, for performing in-memory computations. Moreover, we go a step further and propose the novel `read-compute-store' scheme, where the computed result can be stored \textit{in-situ}, within the memory array, without the need for latching the result and performing a subsequent memory-write instruction. In addition, recently memristor like multi-bit dot product computations using 8T cells has been proposed in \cite{jaiswal20188t}. The present works differs from the work in \cite{jaiswal20188t} since the computations presented in \cite{jaiswal20188t} are analog-like computations in SRAM arrays and requires more complex peripheral circuitry, whereas the focus of the present work is purely digital vector computations in the SRAM arrays.

In addition, almost all beyond CMOS non-volatile technologies have been extensively explored for possible applications to in-memory computing \cite{naturenvm}. These include works based on resistive RAMs \cite{rramimc}, spin-based magnetic RAMs \cite{jain2,kang20,Lee_2013}, and phase change materials \cite{Sebastian_2017}. Such emerging non-volatile technologies promise denser integration, energy-efficient operations and non-volatility as compared to the CMOS based memories, and are suitable for in-memory computations \cite{pinatubo}. However, these emerging technologies are still under extensive research and development phase and their large scale commercialization for on-chip memories is far-fetched.

In this work, we explore in-memory vector operations in \textit{standard} CMOS 8T- and 8$^+$T Differential SRAM cells with minimal modifications in the peripheral circuitry. We call the augmented version of the SRAM bit-cells with extra in-memory compute features as \textit{the X-SRAM}. We propose \textit{at least six different techniques} to enable Boolean computations. The 8T and 8$^+$T cells lend themselves easily for enabling in-memory computations because of the following three factors. 1) The read ports of the 8T and 8$^+$T cells are isolated and can be easily configured to enable in-memory operations. 
%We would, in fact, show that the usual read operation of the standard 8T cell can be exploited to construct NAND and NOR Boolean operations along with the XOR functionality. Further, we would also demonstrate that by applying appropriate voltages, the read-ports of two selected 8T bit-cells can be configured as a voltage divider, which leads to in-memory IMP (implication) and XOR gates. 
2) Also, in sharp contrast to the 6T cells, 8T and 8$^+$T cells do not suffer from read disturb and hence multiple read word-lines within the memory array can be simultaneously activated. 3) In addition, in this manuscript, we exploit the two port structure of the 8T and 8$^+$T cells to propose a novel \textit{read-compute-store} operation, wherein, the computed Boolean data can be stored into the memory array without actually latching the data followed by a subsequent memory write-operation. Later in Appendix, we describe the in-memory computations in standard 6T-SRAMs using the staggered activation of word-lines, as was presented for analog computing in Ref. \cite{shanbhag}.

Some of the key highlights of the present work in comparison to previous works are enumerated below. 
\begin{enumerate}

%\item Standard 6T-SRAM: We propose a sequentially pulsed WL technique for performing a class of bit-wise Boolean logic operations, like NAND/NOR/XOR, using two asymmetric differential current sense amplifiers (SA). The usual memory read/write functionality of the SRAM cell is not disturbed. We also show that the same hardware, including the SA, can be shared for an in-memory operation and a normal memory read operation. Moreover, we show that the extra hardware enhances the memory read operation, by acing as a check for read failures. 

\item We firstly leverage the fact that two simultaneously activated read-word-lines for the standard 8T cells are inherently `wire NORed' through the read bit-line. By using a skewed inverter at the sensing output, we demonstrate that NOR operation can be easily achieved. Further, we also show that NAND logic can similarly be accomplished using another skewed inverter. Note, unlike 6T cells, simultaneous activations of two read word-lines do not impose any read-disturb concerns, thereby opening up a wider design space for optimization.

\item Further, by applying appropriate voltages, we show that two activated read ports of the 8T cell can be configured as a voltage divider. Based on such \textit{voltage divider scheme} we present in-memory vector IMP as well as XOR logic gates. The voltage divider scheme not only allows in-memory computations, but also augments the read mechanism by allowing a possible two bit-read operation under specific conditions.

\item Subsequently, we also present in-memory NAND and NOR computations (along with XOR) in the recently proposed 8$^+$T cells \cite{jp8tdiff}, using \textit{asymmetric sense amplifiers} (SA). The 8$^+$T cells are more robust since they allow differential read sensing as opposed to the standard 8T cells that are characterized by single ended sensing. The usual memory read/write functionality of the SRAM cell is not disturbed due to the use of asymmetric sense amplifiers. We also show that the same hardware, including the SA, can be shared for an in-memory operation and also for the normal memory read operation. Moreover, the extra hardware enhances the memory read operation, by acting as a check for read failures.

\item We propose a novel \textit{`read-compute-store'} scheme for the 8T and 8$^+$T bit-cells, wherein the computed data can directly be written into the desired memory location, without having to latch the output and perform a subsequent memory write operation. This exploits the decoupled read-write paths of the 8T and 8$^+$T bit-cells.

\item We perform Monte-Carlo simulations including voltage and temperature variations to verify the robustness of the proposed in-memory operations for the 8T and the 8$^+$T bit-cells. Energy, delay and area numbers have been presented for each of the proposed scheme.

\item We demonstrate the effectiveness of using in-memory bitwise computations in a typical von-Neumann machine, wherein the conventional SRAM is replaced by the proposed X-SRAM for Advanced Encryption Standard (AES) algorithm. Our system level simulations indicates 75\% reduction in memory accesses thereby saving energy expensive data transfers. 

\end{enumerate}

\begin{figure*}[t]
\centering
\includegraphics[width=0.9\textwidth]{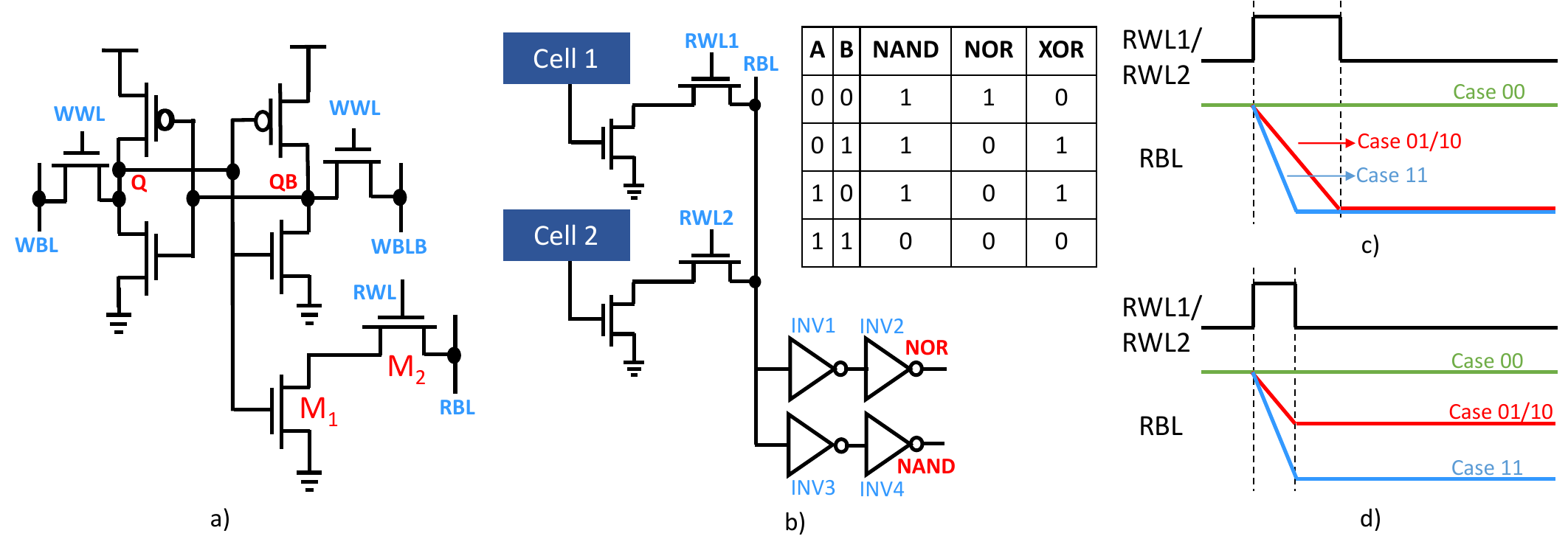}
\caption{a) Schematic of a standard 8T-SRAM bit-cell. In addition to the standard 6T cell, two additional transistors form the read path using a separate read bit-line (RBL). b) Single ended sensing of NAND/NOR using gated skewed inverters. Figure also shows the truth table for NAND/NOR/XOR operations. c) Timing diagram for reading NOR output of Cell 1 and Cell 2. d) Timing diagram for reading NAND output of Cell 1 and Cell 2. }
\label{8t_scheme}
\end{figure*}

\section{In-Memory Computations in 8-Transistor SRAM Bit-Cells}
\label{sec:8t}
As discussed in the introduction, 8T cells have favorable bit-cell structure to enable in-memory computing. Specifically we would exploit the isolated read mechanism and the two port cell topology to embed NAND, NOR, IMP and XOR logic within the memory array. Further, by leveraging the separate read and write ports of the 8T cell, we also propose a \textit{`read-compute-store'} scheme, wherein, by minimal changes in the peripheral circuits, the computed Boolean results can be stored in the desired row of the memory array in the same cycle without the need of latching the results and performing a subsequent write operation.

For each proposal, we first describe the circuit operation using representative illustrations of the transient waveforms followed by actual SPICE based transient simulations under Monte-Carlo analysis. Further, we also present a distribution graph for the key voltages that represent worst case scenarios including temperature as well as voltage variations. Note, in general global process variations can be taken care by proper calibrations, therefore, we concentrate on intra-die threshold voltage variation along with variations in temperature and supply voltage. Towards the end of the manuscript, we tabulate the pros-and-cons of the proposed techniques in a comparative manner.

\begin{figure*}[!t]
\centering
\includegraphics[width=0.9\textwidth]{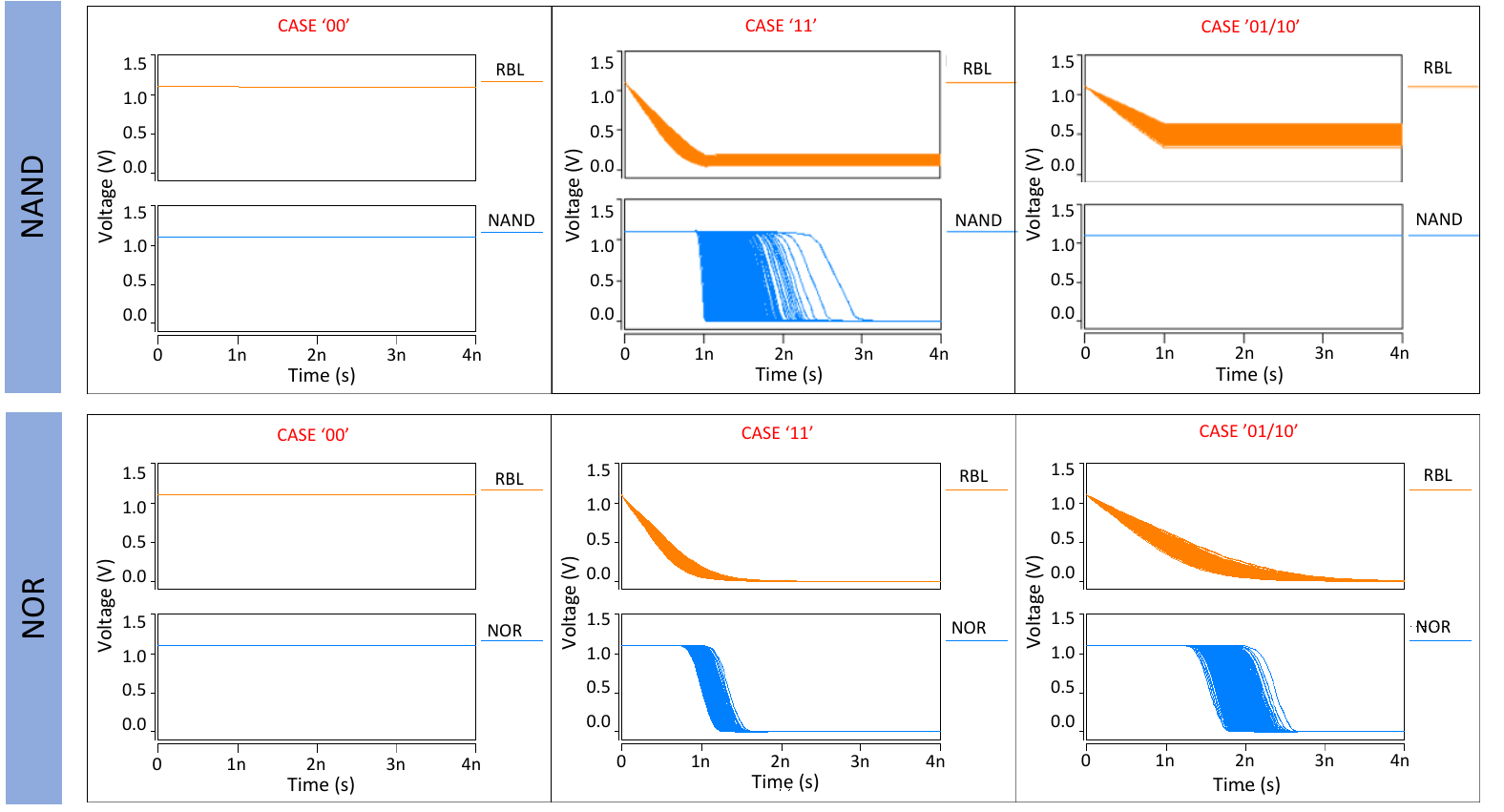}
\caption{Monte-Carlo simulations in SPICE for NAND and NOR outputs for all possible input cases $-$ `00,01,10,11', in presence of 30mV sigma variations in threshold voltage.}
\label{8t_mc}
\end{figure*}

\begin{figure*}[!t]
\centering
\includegraphics[width=0.9\textwidth]{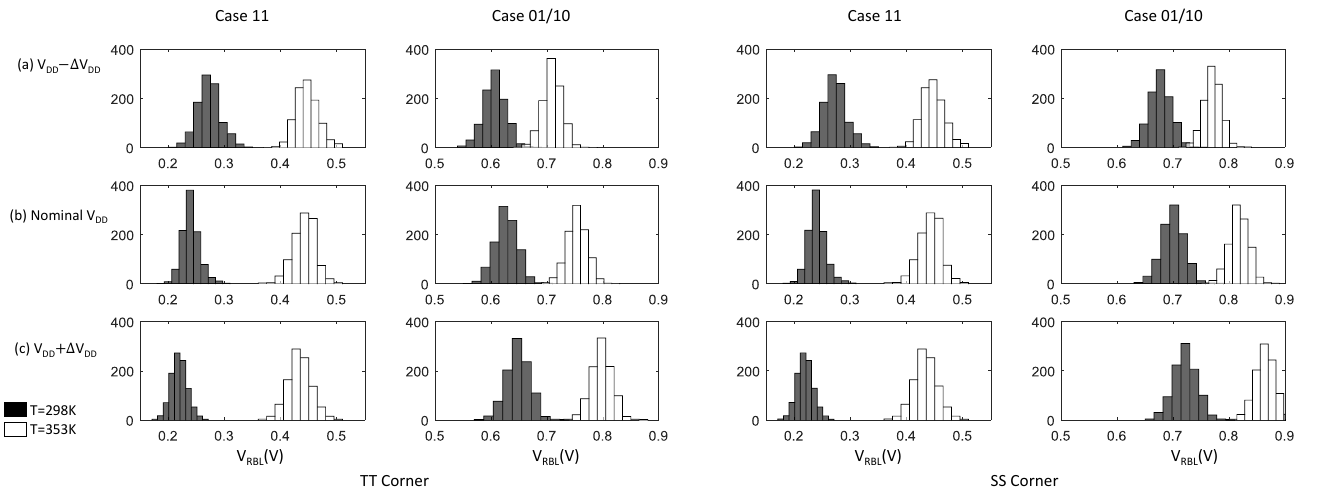}
\caption{Monte-Carlo simulations across process corners (TT corner and SS corner shown) under voltage and temperature variations for NAND outputs for the borderline cases $-$ `01/10' and `11'. The distribution of RBL voltage is plotted under 30mV sigma threshold voltage variations for two different temperatures and $\pm10\%$ variation in nominal $V_{DD}$. }
\label{pvt_8t}
\end{figure*}

\subsection{8-Transistor SRAM: NOR operation}

The 8T SRAM cell is shown in Fig. \ref{8t_scheme}(a). It consists of the usual 6T cell augmented by additional read port constituted by transistors M1-M2. The write operation is similar to the 6T cell, whereas for the read operation, RWL is activated (WWL is low). The RBL is initially pre-charged and if Q = `1' the RBL discharges otherwise it stays at its initial precharged condition. This decoupled read port for the 8T cell allows to have large voltage swing (almost rail-to-rail) on the RBL during the read operation without any concerns of read disturb failure.

The output of a NOR operation is `1' only if both the inputs are `0'. %For the memory implementation it implies that only if both the bits corresponding to the operands `A' and `B' store `0', the logic output should be `1'. In all other cases the output should remain `0'. 
Consider we activate two RWLs corresponding to the rows storing vector operand `A' and vector operand `B', respectively, as shown in Fig. \ref{8t_scheme}(b). Due to the decoupled read ports, both the RWLs can be activated simultaneously without any read disturb concerns as opposed to the 6T cell. The precharged RBL line retains its precharged state if and only if both the bits Q corresponding to operands `A' and `B' are `0'. In other words, as shown in Fig. \ref{8t_scheme}(c) RBL remains high only if Q = `0' for both `A' and `B'. Thus, merely by activating the two RWLs, data stored in the two bit-cells are `wire NORed'. A gated inverter (INV1) is connected to the RBL such that the inverter output goes low if the RBL remains high. Thereby, the output of the cascaded inverter (INV2) mimics the NOR operation. Note, the NOR operation is same as the usual read operation except that we have turned ON two RWLs instead of one. Thus, NOR can be easily achieved in the 8T bit-cell without any significant overhead. The timing diagram for the NOR operation is shown in Fig. \ref{8t_scheme}(c). It is also interesting to observe that although we have discussed the NOR operation for two inputs, the proposed scheme can in fact be extended to \textit{n-input} NOR operations. For an \textit{n-input} NOR operation n-read world-lines can be simultaneously activated and RBL would remain high only if all the corresponding operands are `0' which would represent the \textit{n-input} NOR truth table.

\subsection{8-Transistor SRAM: NAND operation}

Let us consider that we activate two RWLs corresponding to vector operands `A' and `B', respectively. The precharged RBL will eventually go to 0V if Q for any one of the input operand is `1'. However, the fall time of the signal at RBL from the precharged value to 0V would depend strongly on the fact, whether any one Q is high or if both the Q bits are high simultaneously. In other words, only if both the Qs are `1', the discharge of the precharged RBL line would be fast enough. In Fig. \ref{8t_scheme}(d), we have shown schematically the state of the RBL for all input cases. In order to exploit the different discharge rates of the RBL, the RWL signal had to be timed such that the RBL does not discharge completely in cases `01/10'. This allows a difference in voltage levels on RBL in the two cases (`01/10' and `11'). The trip point of the inverter INV3 is chosen such that it goes high only for the case `11', thus output of inverter INV4 mimics the NAND operation.

Fig. \ref{8t_mc} shows the SPICE transient simulation for the NAND and NOR proposals, under 30mV sigma threshold voltage variation in transistors. We used 45-nm Predictive Technology Models (PTM) \cite{ptm} for simulating the circuits. A BL and BLB capacitance of 10fF was assumed for all the simulations. As discussed earlier, the NAND computation has a narrower design margin due to its timing critical operation as opposed to the NOR logic. Specifically, for the NAND operation a discharge path with two parallel transistors needs to be distinguished from the discharge path with one transistor. To analyze the robustness and the design margin, we performed a rigorous variation analysis across process corners including voltage and temperature variations for the NAND operation, as shown in Fig. \ref{pvt_8t}. Monte-Carlo simulations with 30mV sigma threshold voltage variation were performed along with a $\pm10\%$  variation in nominal $V_{DD}$ ($\Delta V_{DD}$). The simulations were repeated for two different temperatures. The figure shows the resultant distribution of voltage on RBL for the borderline cases `11' and `01/10', at the instant when RWL is pulled LOW. 

In order to study the effect of variations due to different process corners, we also performed simulations assuming global variations in the threshold voltage, the simulations were performed for all possible corners including SS (slow NMOS, Slow PMOS), SF (Slow NMOS, Fast PMOS), FS (Fast NMOS, Slow PMOS) and FF (Fast NMOS, Fast PMOS). The threshold voltages for respective corners were globally shifted in appropriate directions for each of the process corners for both the PMOS and the NMOS transistors. For example, for the SS and FF corners, the threshold voltages were increased or decreased by $\sim$90mV to imitate the affect of process corners. These global shifts in threshold voltages were then super-imposed by random VT variations to evaluate the cumulative effect. In Fig. 5, we have shown the Monte-Carlo results for two different process corners $-$ the nominal case (TT) and for the SS corner, for two different temperatures including $\pm$ 10\% variation in supply voltage. Note, similar results were obtained for other process corners as well, however to avoid clutter, we have shown two representative results for the process corners. It can be observed, we obtain a 50mV worst cases sense margin in Fig. \ref{pvt_8t}(a) for a $-10\%$ nominal $V_{DD}$. Also, the timing for the NAND operation can be controlled by a digitally programmable delay based control signal for tuning the pulse activation of the RWL \cite{maymandi2005monotonic}. Such a programmable delay path would require a one-time calibration depending on the process corner, for proper functionality. 

In addition to the NAND and NOR operations, by NORing the outputs of the AND (INV3) and the NOR (INV2) gates together, XOR operation can be easily achieved. In summary, we have shown that the very bit-cell topology of the 8T cell can be exploited to accomplish in-memory NOR, NAND and XOR computations. In the next sub-section, we would discuss another proposal for embedding IMP as well as XOR gate within the 8T SRAM array by utilizing the proposed voltage divider scheme.

\subsection{8 Transistor SRAM: Voltage Divider Scheme for IMP and XOR gates}

\begin{figure*}[t]
\centering
\includegraphics[width=0.9\textwidth]{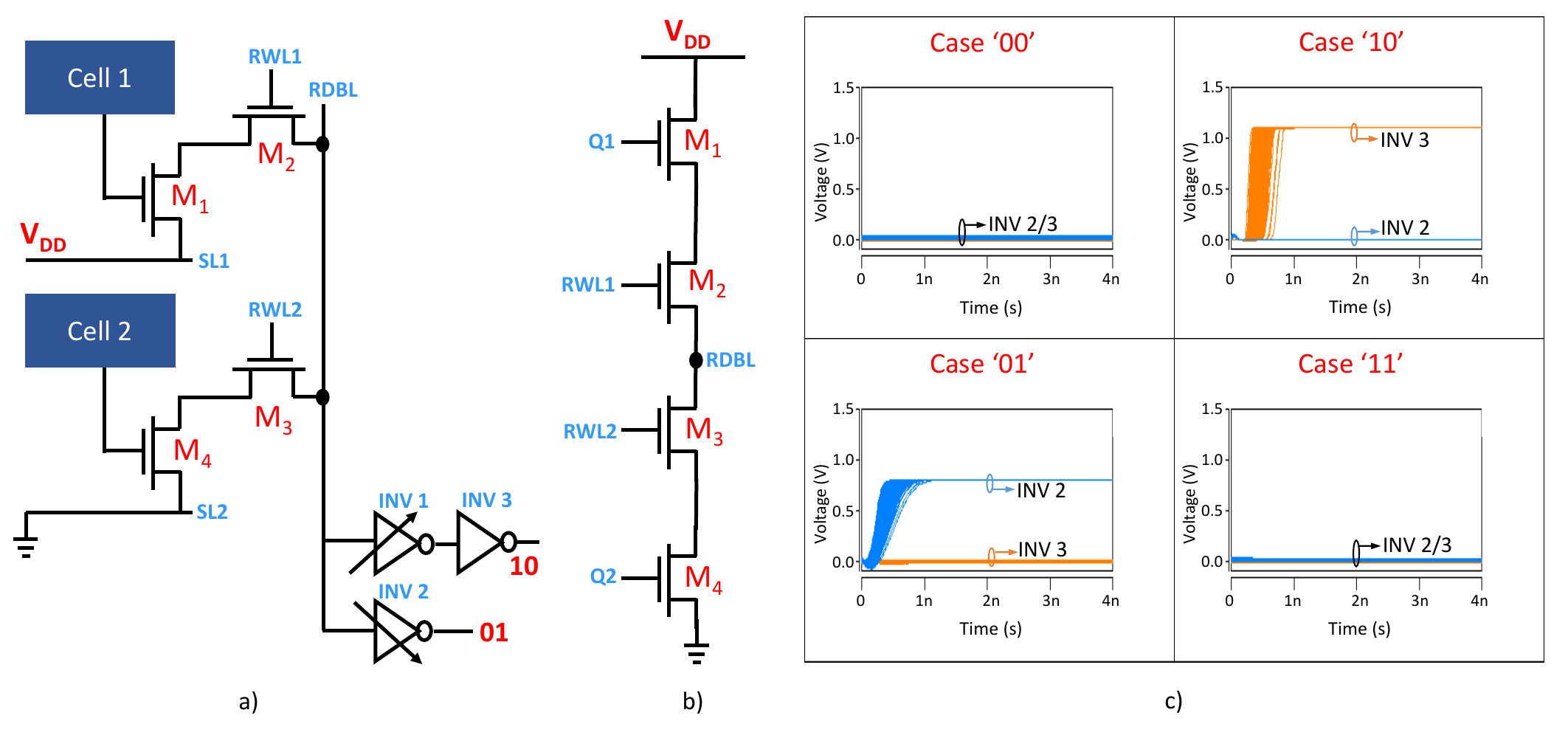}
\caption{a) Circuit schematic of the 8T-SRAM for implementing the voltage-divider scheme. b) Equivalent circuit traced by transistors $M1-M4$ while data is read from Cell 1 and Cell 2. c) Monte-Carlo simulations in SPICE for all possible input cases, showing the output of the two asymmetric inverters.}
\label{8t_appendix}
\end{figure*}

In this sub-section, we present a method of implementing IMP and XOR operation using 8T cell by exploiting the voltage divider principle. Let us consider, the circuit shown in Fig. \ref{8t_appendix}(a). Let us assume the first operand is stored in the upper bit-cell corresponding to the line RWL1, while the second operand is stored in the lower bit-cell corresponding to RWL2. In the conventional 8T cell, the source of transistors M$_1$ and M$_4$ are connected to ground. In the presented circuit, the source of the transistors M$_1$ and M$_4$ are connected to respective source lines (SL1 and SL2 shared along respective rows). During the normal operations, the SLs can be grounded, thereby accomplishing usual 8T SRAM read and write operations. 

During the in-memory computation mode, the SL1 is pulled to V$_{DD}$, while the SL2 is grounded. RWL1 and RWL2 are initially grounded and RDBL is pre-charged to a voltage V$_{pre}$ (chosen to be 400mV). After the pre-charge phase, transistors M$_2$ and M$_4$ are switched ON, thereby M$_1$ $-$ M$_2$ $-$ M$_3$ $-$ M$_4$ form a voltage divider and RDBL forms the middle node of the voltage divider structure (see Fig. \ref{8t_appendix}(b)). Note, in the voltage divider configuration, M$_1$ and M$_2$ are strongly source degenerated. In order to make sure M$_1$ and M$_2$ are sufficiently ON, we boosted the $V_{DD}$ of `Cell 1' and RWL1 such that the gate of M$_1$ and M$_2$ have enough overdrive when the `Cell 1' is storing a digital `1' (Q = `1' and QB = `0').

\begin{figure*}[t]
\centering
\includegraphics[width=\textwidth]{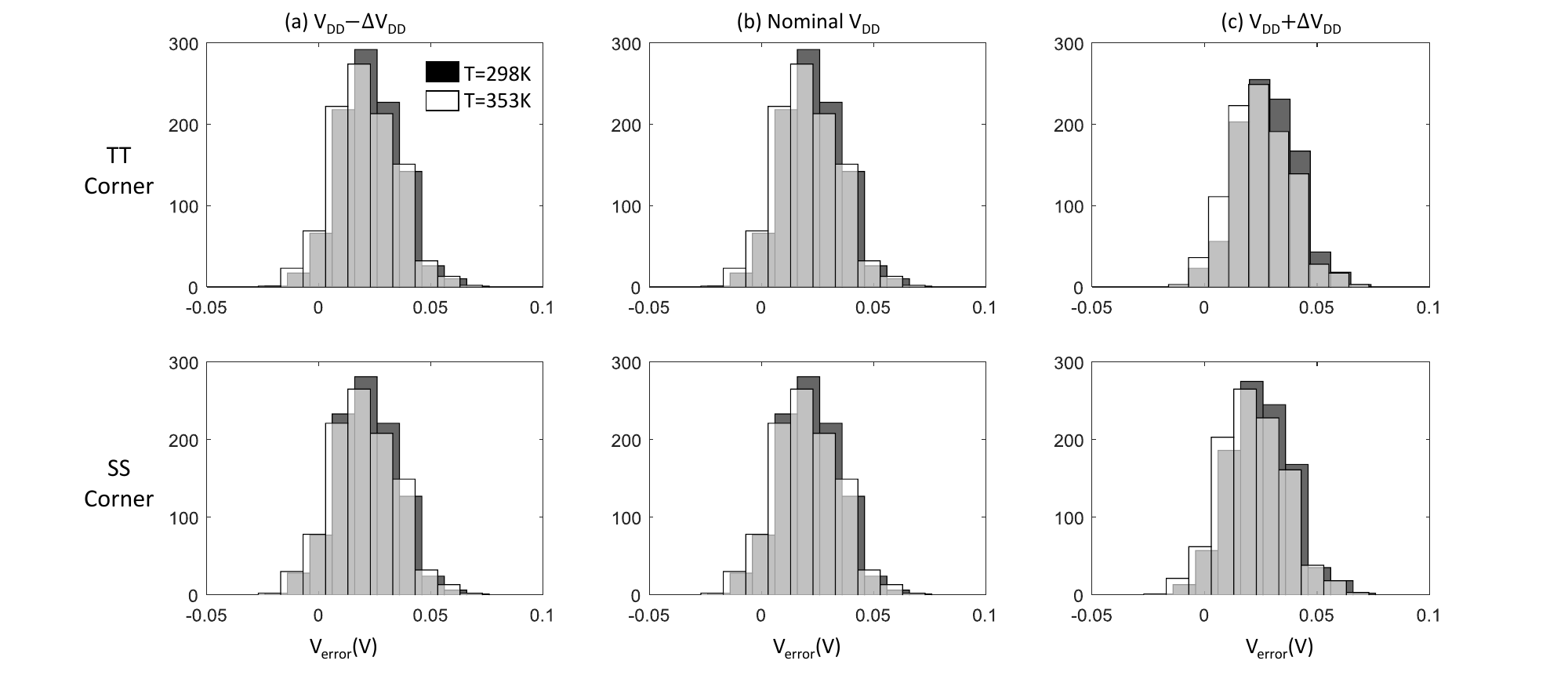}
\caption{Monte-Carlo simulations with variations in supply voltage and temperature across process corners for the voltage-divider scheme for the case (1,1). $V_{error}$ is defined as the difference between the RBL voltage (when both the operands are `1') and the initial pre-charge voltage V$_{pre}$. The distribution of $V_{error}$ is plotted under 30mV sigma threshold voltage variations for two different temperatures and $\pm10\%$ variation in nominal $V_{DD}$. The variations in V$_{pre}$ are also accounted for.}
\label{pvt_vdiv}
\end{figure*}

In the voltage divider configuration M$_1$ $-$ M$_2$ $-$ M$_3$ $-$ M$_4$, RDBL retains its precharged voltage V$_{pre}$ if both the bit-cells are storing digital `0' (\textit{i.e.} M$_1$ and M$_4$ are OFF ). Similarly, if both the cells are storing a digital `1' (\textit{i.e.} M$_1$ and M$_4$ are ON), the voltage at RDBL stays close to its precharged value (400mV) due to the voltage divider effect. Thus, when the cells store (0,0) or (1,1) (where the first (second) number in the bracket indicates the data stored in Cell 1 (2)), the voltage at RDBL stays close to the precharged voltage. On the other hand, if the data stored is (1,0), then M$_1$ is ON while M$_4$ is OFF. As such, RDBL will charge to V$_{DD}$ through transistors M$_1$ and M$_2$. In contrast, if the data stored is (0,1), M$_4$ is ON while M$_1$ is OFF. Therefore, RDBL will discharge to 0V through transistors M$_3$ and M$_4$. In summary, the voltage on RDBL stays close to V$_{pre}$ when both the cells store same data. RDBL charges to V$_{DD}$ for data (1,0) and discharges to 0V for data (0,1).

The state of the data stored in the two cells can be sensed through two skewed inverters. INV2 is skewed such that it goes high only when RDBL is much lower than V$_{pre}$ and is close to 0V, while INV1 is skewed so that it goes low only when RDBL is higher than V$_{pre}$ and is close to V$_{DD}$. In other words, high output at INV2 indicates data (0,1) while high output at INV3 indicates data (1,0). Interestingly, INV1 implements `A IMP B'. By ORing the output of INV2 and INV3 we can obtain the XOR of inputs A and B. 

Fig. \ref{pvt_vdiv} shows the distribution of $V_{error}$ (defined below) under $V_T$ variations in addition to variations in temperature and supply-voltage. Monte-Carlo simulations across process corners, similar to the ones performed for NAND in the previous sub-section were performed in this case. Note, when either of the two operands Q1 or Q1 is low, the circuit in Fig. \ref{8t_appendix}(b) reduces to an RC charging or discharging circuit, respectively. As such, if any of Q1 or Q2 is low, the RDBL would either charge up to $V_{DD}$ or discharge to ground even under variations. The critical case arises when both Q1 and Q2 are low or high, simultaneously. For robust operation, ideally we want the RDBL voltage to stay at $V_{pre}$ for both the cases (Q1 = Q2 = low or Q1 = Q2 = high). Therefore, we analyze the difference in voltages on the RDBL in the two cases `Q1 = Q2 = low' versus `Q1 = Q2 = high'. We define $V_{error}$ as the difference between the RDBL voltage for case (1,1) and case (0,0). In other words, $V_{error}$ denotes the variation of RDBL voltage when the voltage divider is active with respect to V$_{pre}$. The variations in V$_{pre}$ are also considered in the Monte-Carlo simulations. We observe that $V_{error}$ is close to zero, making this configuration robust to variations, as shown in Fig. 7. Intuitively, the robustness of the proposed scheme stems from the fact that changes in voltage and temperature affects all the four transistors of Fig. \ref{8t_appendix}(b) in similar manner thereby reducing any variations in the voltage at node RDBL. Moreover, since we use the static voltage developed at RDBL, unlike the time-sensitive discharge in the earlier scheme, the voltage-divider scheme is robust to process corners as well. The voltage at RDBL depends on the relative strengths of the four transistors. Since process corners induce global $V_T$ shifts, all NMOS transistors are equally affected, making the voltage-divider ratio largely unaffected. This is evident form the two representative process corner simulations shown in Fig. 7.

Some key features of the voltage divider logic scheme are, 1) IMP is a universal gate and hence any arbitrary Boolean function can be implemented using the proposed scheme 2) if any one of the inverter outputs (INV2 or INV3) are high, it indicates the data stored is (0,1) or (1,0), thereby allowing a two bit-read operation in addition to the desired in-memory computation. However, if none of the inverters are high then a subsequent read operation would be required to ascertain if the stored data is (0,0) or (1,1). As such, in 50\% cases when the data stored is (0,1) or (1,0), we can accomplish a two bit read operation, along with the in-memory compute operation.

\subsection{Proposed `read-compute-store' (RCS) scheme}

We have seen that basic Boolean operations like NAND, NOR, IMP and XOR can be computed using 8T cells. We would now show that the decoupled read and write ports of the 8T bit-cell can be used for enabling `read-compute-store' (RCS) scheme. The RCS scheme implies that while the data is being read from the two activated RWLs (corresponding to the two input operands), simultaneously the WWL of a third row can be activated such that the computed data gets stored in the third row at the same time while the actual Boolean computation is in progress. As such, the computed data is not required to be latched first, then written subsequently, in a multi-cycle fashion. Note, writing into 8T bit-cells is much easier due to the fact that the write port of the 8T cell is specifically optimized for the write operation.

\begin{figure*}[!t]
\centering
\includegraphics[width=0.75\textwidth]{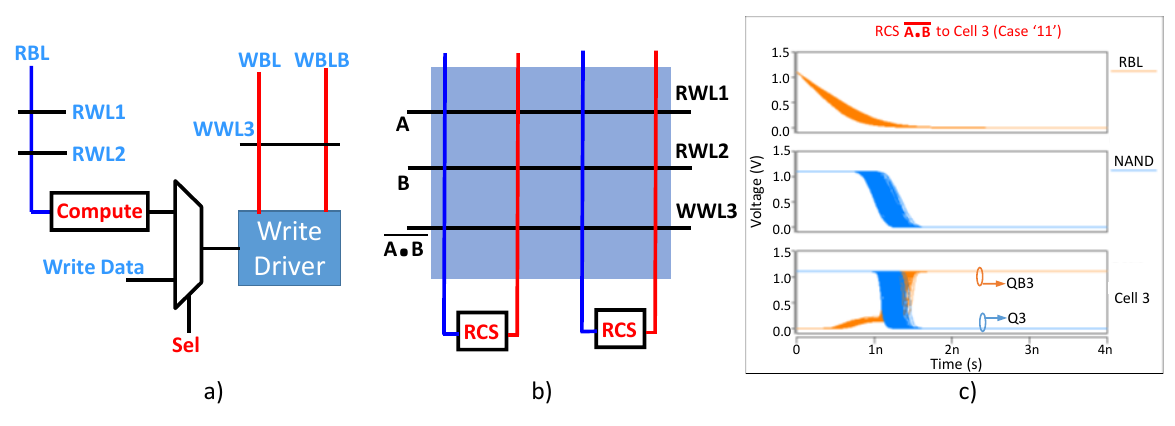}
\caption{a) Proposed `read-compute-store' (RCS) scheme. RWL1 and RWL2 are enabled, corresponding to the data to be computed. The computation output is selectively passed to the write-driver of that column, while simultaneously enabling the WWL3, where data is to be stored. b) Block diagram showing the RCS blocks in the memory array. The NAND of row 1 and row 2 is to be stored in row 3. c) Monte-Carlo simulations in SPICE, showing the final state of Cell 3 stores the desired output.}
\label{RCS}
\end{figure*}

Let us understand how the RCS scheme can be implemented with reference to Fig. \ref{RCS}. Assume that the input operands correspond to the rows 1 and 2, while the resulting Boolean computation has to be stored in row 3. Note, this Boolean computation can be either of NAND/NOR/IMP/XOR. Let us take the example for the NAND operation. As shown in Fig. \ref{RCS}(a), two read lines RWL1 and RWL2 would be activated, the compute block, which basically is the abstracted view of the skewed inverters of Fig. \ref{8t_scheme}(b), would perform the logic computation. Now, since the read and write port for 8T cell are decoupled we can simultaneously activate a third WL, in this case the write word-line (WWL3). The computed output can be selected through a multiplexer and fed to the write drivers for directly storing the Boolean result in the bit-cells corresponding to WWL3. Thus, the fact that 8T cells have decoupled read-write ports can be leveraged to accomplish the proposed `read-compute-store' scheme. Fig. \ref{RCS}(b) shows schematically the array level block diagram where the three word-lines RWL1, RWL2 and WWL3 are activated simultaneously. In Fig. \ref{RCS}(c) we show the Monte-Carlo results for storing the computed NAND output into Cell3. Note that a `copy' operation can also be performed using the RCS scheme, by activating the RWL of the source row and WWL of the destination row. In this case, the input to the RCS block will simply be the SA output, which corresponds to the data stored in the bit-cells of the source row. 

%In addition, a \textit{voltage divider scheme} can be used in conjunction to the 8T cell for computing implication (IMP) and XOR logic. However, this scheme would require voltage boosting and hence has been described in the appendix. Nevertheless, the voltage divider scheme allows yet another method of enabling Boolean computations in the 8T cells. The RCS scheme is equally valid for the voltage divider scheme described in the appendix.

\section{8$^+$ Transistor Differential Read SRAM}

\begin{figure*}[t]
\centering
\includegraphics[width=0.8\textwidth]{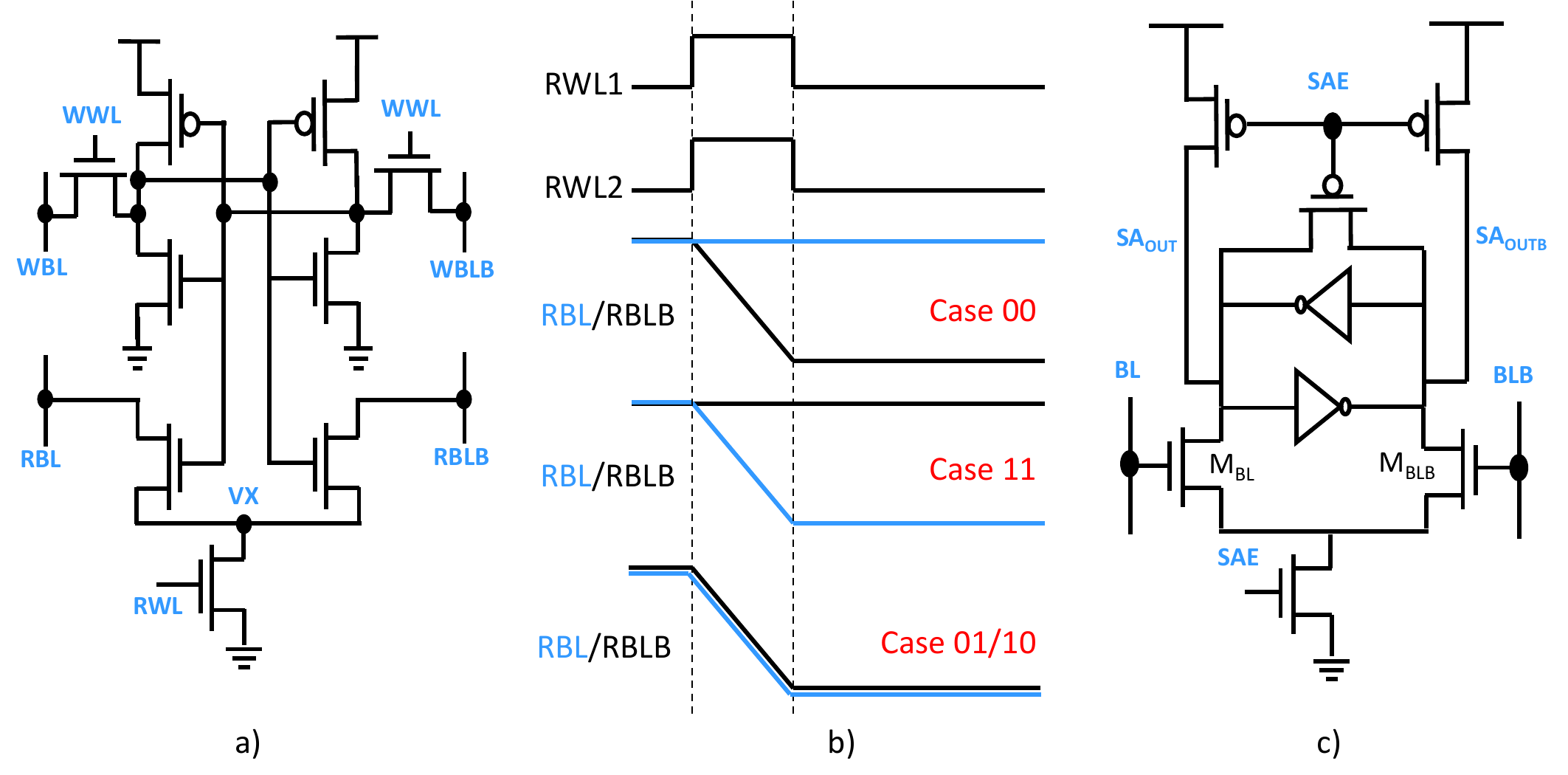}
\caption{a) Circuit schematic of an 8$^+$T Differential SRAM bit-cell \cite{jp8tdiff}. b) Timing diagram used for in-memory computations on the 8$^+$T Differential SRAM. c) Circuit schematic of the proposed asymmetric differential sense amplifier.}
\label{fig:diff8t}
\end{figure*}

Recently, an 8$^+$T Differential SRAM design was proposed in \cite{jp8tdiff} to overcome the single ended sensing of the conventional 8T-SRAM cell. 8$^+$T Differential SRAM has decoupled read-write paths with an added advantage of a differential read mechanism through the read bit-lines RBL/RBLB (see Fig. \ref{fig:diff8t}(a)), as opposed to the single-ended read mechanism of 8T-SRAM. The ninth transistor, whose gate is connected to RWL in Fig. \ref{fig:diff8t}(a) is shared by all the bit cells in the same row. The differential read operation is very similar to the read operation of a standard 6T-SRAM. The usual memory read operation is performed by pre-charging the bit-lines (RBL and RBLB) to V$_{DD}$, and subsequently enabling the word-line corresponding to the row to be read out. Depending on whether the bit-cell stores `1' or `0', RBL or RBLB discharges. The difference in voltages on RBL and RBLB is sensed using a differential sense amplifier.

Let us consider words `A' and `B' stored in two rows of the memory array. Note that we can simultaneously enable the two corresponding RWLs without worrying about read-disturbs, since the bit-cell has decoupled read-write paths. The RBL/RBLB are pre-charged to V$_{DD}$. For the case `AB'=`00' (`11'), RBL (RBLB) discharges to 0V, but RBLB (RBL) remains in the precharged state. However, for cases `10' and `01', both RBL and RBLB discharge simultaneously. The four cases are summarized in Fig. \ref{fig:diff8t}(b).

\begin{figure*}[t]
\centering
\includegraphics[width=0.85\textwidth]{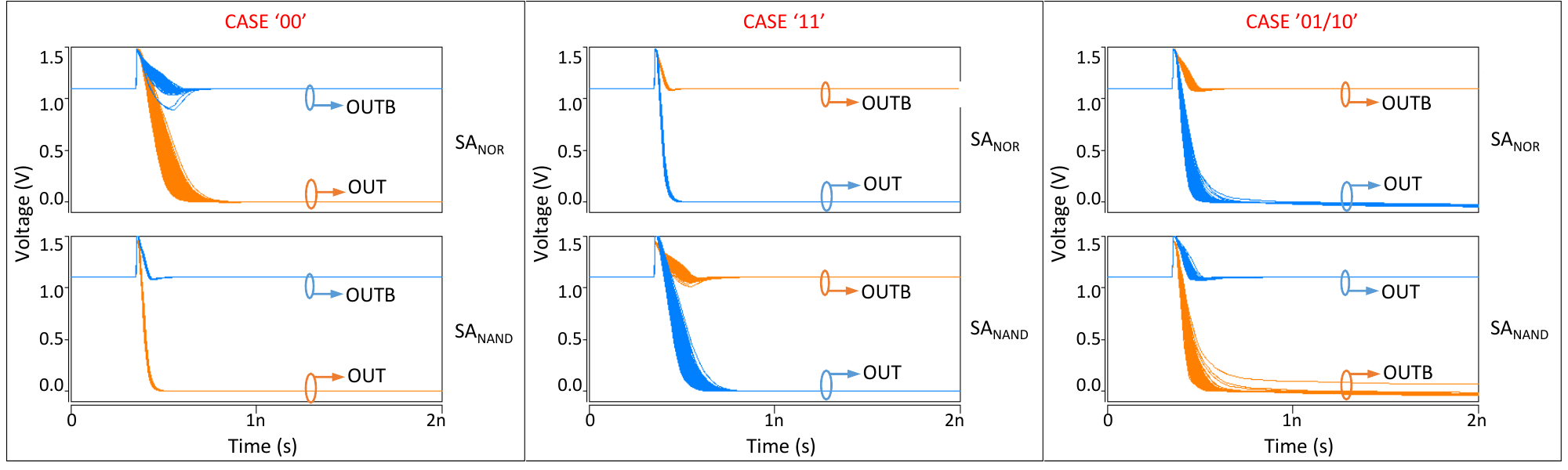}
\caption{Monte-Carlo simulations in SPICE for SA outputs for all possible input cases $-$ `00,01,10,11', in presence of 30mV sigma variations in threshold voltage.}
\label{8tdiff_mc}
\end{figure*}

\begin{figure*}[t]
\centering
\includegraphics[width=0.9\textwidth]{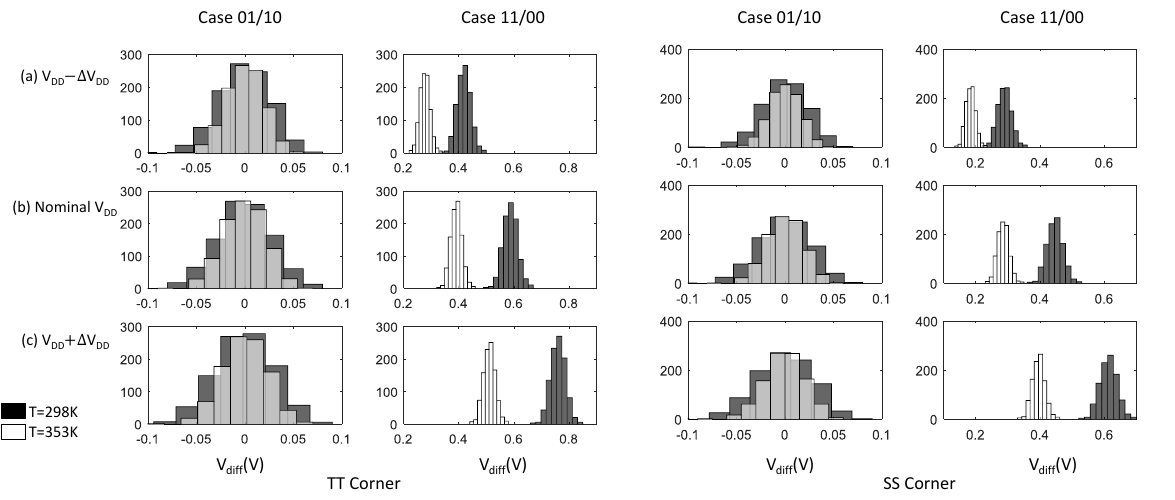}
\caption{Monte-Carlo simulations across process corners under $V_T$ and temperature and supply-voltage variations for the 8$^+$T SRAM configuration for the cases `01/10' and `11/00'. $V_{diff}$ is defined as the absolute difference between the RBL and RBLB voltages at the instant when the sense amplifier is enabled. The distribution of $V_{diff}$ is plotted under 30mV sigma threshold voltage variations for two different temperatures and $\pm10\%$ variation in nominal $V_{DD}$.}
\label{pvt_8tdiff}
\end{figure*}

Now, in order to sense bit-wise NAND and NOR operation of `A' and `B', we propose an asymmetric SA (see Fig. \ref{fig:diff8t}(c)), by skewing one of the transistors. Skewing the transistors can be done in multiple ways, for example, transistor sizing, threshold voltage, body bias \textit{etc}. In Fig. \ref{fig:diff8t}(c), if the transistor $M_{BL}$ is deliberately sized bigger compared to $M_{BLB}$, its current carrying capability increases. For cases `01' and `10', both RBL and RBLB discharge simultaneously. However, since the current carrying capability of $M_{BL}$ is more than $M_{BLB}$, $SA_{out}$ node discharges faster, and the cross-coupled inverter pair of the SA stabilizes with $SA_{out}$=`0'. For the case `11', RBL starts to discharge, while RBLB is at V$_{DD}$. The SA amplifies the voltage difference between RBL and RBLB, resulting in $SA_{out}$=`1'. Whereas for the case `00', RBLB starts to discharge, while RBL is at V$_{DD}$, giving $SA_{out}$=`0'.

\begin{table*}
\centering
\includegraphics[width=\textwidth]{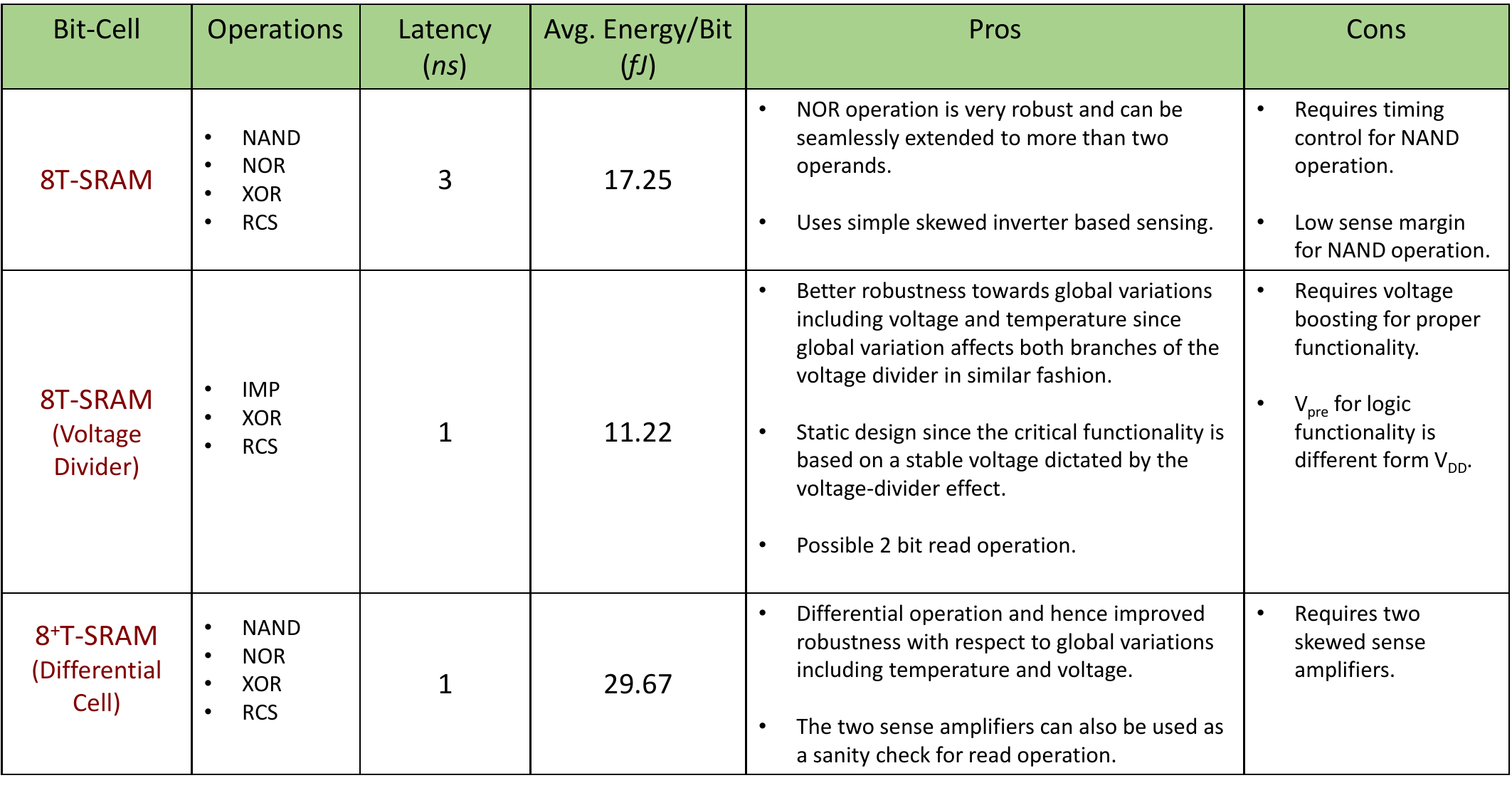}
\caption{Summary of proposals described in the manuscript. The table shows average energy consumption per-bit and latency for the in-memory operations on various bit-cells. Pros and cons of each proposal are also listed.}
\label{table:par}
\end{table*}

%\begin{table}[t]
%\renewcommand{\arraystretch}{1.3}
%\centering
%\caption{Average energy per-bit and latency for the proposed in-memory operations on various bit-cells.}
%\label{table:par}
%\begin{tabular}{|c|c|c|}
%\hline
% \bfseries{Bit-Cell} &   \bfseries{Latency (ns)} &   \bfseries{Average Energy/bit (fJ)} \\
%\hline
%8T-SRAM & 3 & 17.25 \\
%8T-SRAM (Voltage Divider) & 1 & 11.22 \\
%8$^+$T Differential SRAM & 1 & 29.67 \\
%6T-SRAM (Appendix) & 3 & 29.3\\
% \hline
% & {Shared read-write paths}  & {Separate read-write paths, single ended sensing} & {Separate read-write paths, differential sensing}  \\
% Comments   & {Prone to read disturb failure}  &{No read disturb concerns} & {No read disturb concerns}  \\
% & {Uses Asymmetric SA}  &{Supports `read-compute-store'} & {Supports `read-compute-store'}  \\
% & {}  &{Voltage-divider scheme for IMP and XOR (Appendix)} & {}  \\
%\hline \hline
%\end{tabular}
%\end{table}

Thus it can be observed that $SA_{out}$ generates an AND gate (thus, $SA_{outb}$ outputs NAND gate). Similarly, by sizing the $M_{BLB}$ bigger than $M_{BL}$, OR/NOR gates can be obtained at the SA outputs. Finally, two SAs in parallel (one with $M_{BL}$ up-sized, $SA_{NAND}$, and one with $M_{BLB}$ up-sized, $SA_{NOR}$) enable bit-wise AND/NAND and OR/NOR logic gates. Moreover, an XOR gate can be obtained by combining the AND/NAND and OR/NOR outputs using an additional NOR gate. Thus, in a single memory read cycle, we obtain a class of Boolean bitwise operations, read directly from the asymmetrically sized SA outputs. SPICE transient simulations with 30mV sigma variations in the threshold voltage for all input data cases are summarized in Fig. \ref{8tdiff_mc}. Monte-carlo analysis across process corners with $V_T$ variations and variations in supply voltage and temperature for the 8$^+$T Differential SRAM are shown in Fig. \ref{pvt_8tdiff}. $V_{diff}$ is defined as the absolute difference between the RBL and RBLB voltages at the instant when the sense amplifier is enabled. For the case `01/10', $V_{diff}$ should be close to 0V to allow the asymmetry in the SA to determine the output. Whereas, for the case `11/00', $V_{diff}$ should be large enough, so that the output is driven by the differential voltage difference between RBL and RBLB, and not due to the asymmetry in SA. The difference in RBL and RBLB voltages for various cases is shown in Fig. \ref{pvt_8tdiff}. Due to global $V_T$ variations across process corners, the discharge on both RBL/RBLB is affected in the same manner. Moreover, since we use a differential voltage sense-amplifier, these global variations cancel, thereby making this scheme robust to process corners. Note, even in worst case the voltage difference between the `01/10' and `11/00' case is sufficient for proper differential SA operation. Interestingly, this difference also increases with increase in $V_{DD}$, hence voltage boosting can be easily employed to increase the design margin.

It is worthwhile to note that the two SAs can be used for regular memory read operations as well. The two cases of a typical memory read operation are similar to the cases `11' and `00' in Fig. \ref{fig:diff8t}(b). Both SAs will generate the same output corresponding to the bit stored in the cell. Moreover, the output of the XOR gate inherently acts as an in-memory check for possible read failures. The RCS scheme described in Section \ref{sec:8t} can also be applied to 8$^+$T Differential SRAMs due to decoupled read-write paths. Along with the two RWLs from where the input operands are read, a WWL can also be enabled which would eventually store the Boolean output within the memory array in the same cycle.

Using 8$^+$T cells is advantageous over the conventional 8T cells for in-memory bit-wise logic operations because of better robustness due to the differential read operation, in contrast to the single ended read in 8T-SRAM cells.
\label{8tdiff}

\section{Discussions}

\begin{figure*}[t]
\centering
\includegraphics[width=\textwidth]{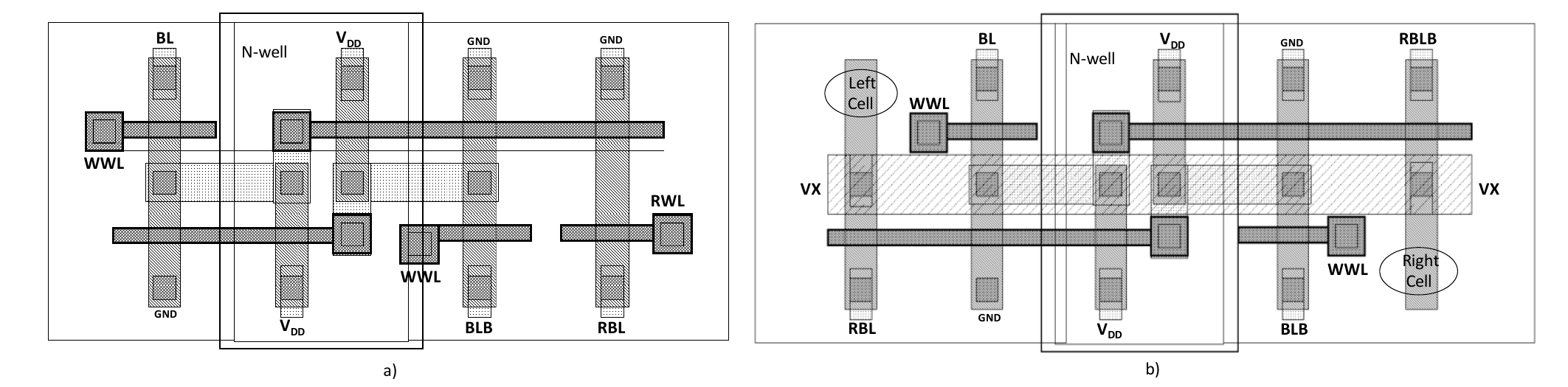}
\caption{a) Thin cell layout for the standard 8T-SRAM bit-cell shown in Fig. \ref{8t_scheme}(a). b) Thin cell layout for the 8$^+$T Differential SRAM bit-cell \cite{jp8tdiff} illustrated in Fig. \ref{fig:diff8t}(a). Left- and right-most diffusion tracks are shared with adjacent bit-cells. The ninth transistor in Fig. \ref{fig:diff8t}(a) is common for the row and is connected at the periphery to the node `VX'.}
\label{fig:layout}
\end{figure*}

In sections II and III, we have seen various ways of implementing basic Boolean operations using the 8T and the 8$^+$T bit-cells. Table \ref{table:par} presents the average energy per-bit and latency for each of the proposed in-memory compute techniques. The 8T cell allows separate read write ports, thereby alleviating any possible read-disturb failure concerns. In addition, it also supports the proposed RCS scheme. However, 8T cell suffers from robustness concerns due to its single ended sensing.

Using the 8$^+$T cell, on the other hand, allows differential sensing like the conventional 6T cell, while also allowing separate read and write ports. It thus combines the benefits of both the standard 6T and the 8T cells. The thin cell layout of standard 8T bit-cells and the 8$^+$T bit-cells are shown in Fig. \ref{fig:layout}. Standard 8T cell requires five diffusion tracks, while the 8$^+$T cell requires six. However, left- and right-most diffusion tracks are shared with adjacent bit-cells, thereby achieving similar area per-bit as compared to the standard 8T cell \cite{jp8tdiff}.

Note, since the differential read scheme for the 8$^+$T cell is functionally similar to the conventional 6T cell, NOR and NAND gates (along with the XOR gate) can also be implemented in the 6T based memory array. However, due to the shared read-write paths of the 6T cell, the word-lines cannot be simultaneously activated and require a sequential activation. In addition, 6T cells are read disturb prone and hence would exhibit much lesser robustness than the proposed 8T and 8$^+$T cells. Nevertheless, in the Appendix we have included a description of how the 6T cells can be used to accomplish NOR, NAND and XOR operations. We also show that an in-memory `copy' operation can also be easily achieved in the 6T cell due to its shared read/write paths.

Finally, it is worth noting that although we have proposed multiple in-memory techniques in this manuscript, the choice of the bit-cell and the associated Boolean function would heavily depend on the target application. Thus, Table \ref{table:par} summarizes the pros and cons of each proposal. The aim of the present manuscript is to demonstrate various possible techniques that can be utilized in conventional CMOS based memories for accomplishing in-memory Boolean computations. Since the present proposal augments the functionality of the memory arrays without changing the basic circuitry, it has wide applications in diverse computing systems, few of them are $-$
1.	A standard von-Neumann general-purpose processor with SRAM replaced by X-SRAM.
2.	A modified GPU, wherein the SRAM based register files are replaces by X-SRAM arrays.
3.	A machine learning or artificial intelligence processor, for example, a binary neural network accelerator.
As an example, in the next section we would present an encryption accelerator using the proposed X-SRAM.

\section{ X-SRAM based non-standard von-Neumann Computing for AES Encryption}

%%%%% NEEDS EDITS %%%%%%%

\begin{figure*}[t]
\centering
\includegraphics[width=\textwidth]{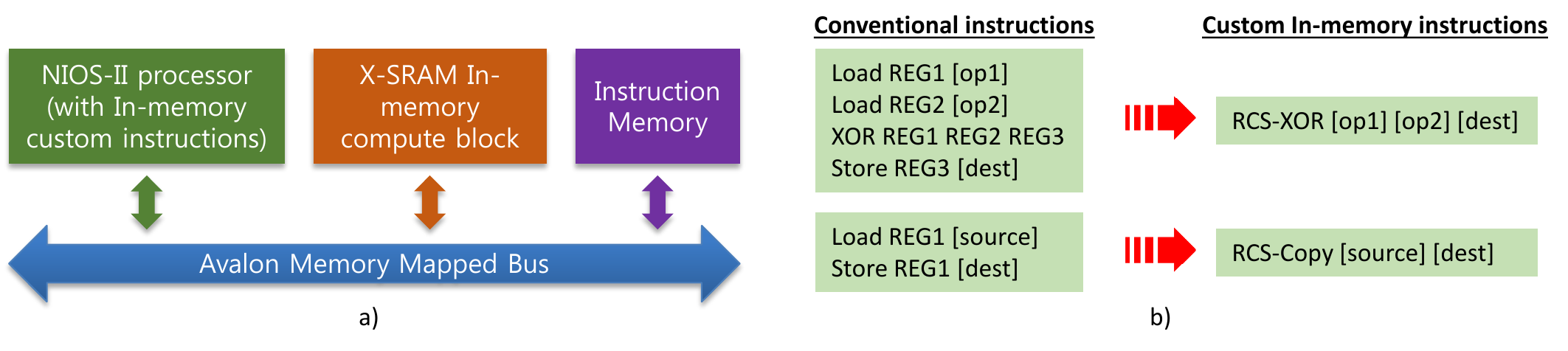}
\caption{(a) System-level implementation of a typical von-Neumann architecture with X-SRAM as the memory block. The processor, data-memory and the instruction-memory blocks are connected via a shared system bus. (b) Illustration of custom in-memory instructions added to the instruction set of the Nios-II processor. Substituting in-memory instructions reduces unnecessary read-writes into the memory.}
\label{systemimpl}
\end{figure*}

\begin{figure*}[t]
\centering
\includegraphics[width=\textwidth]{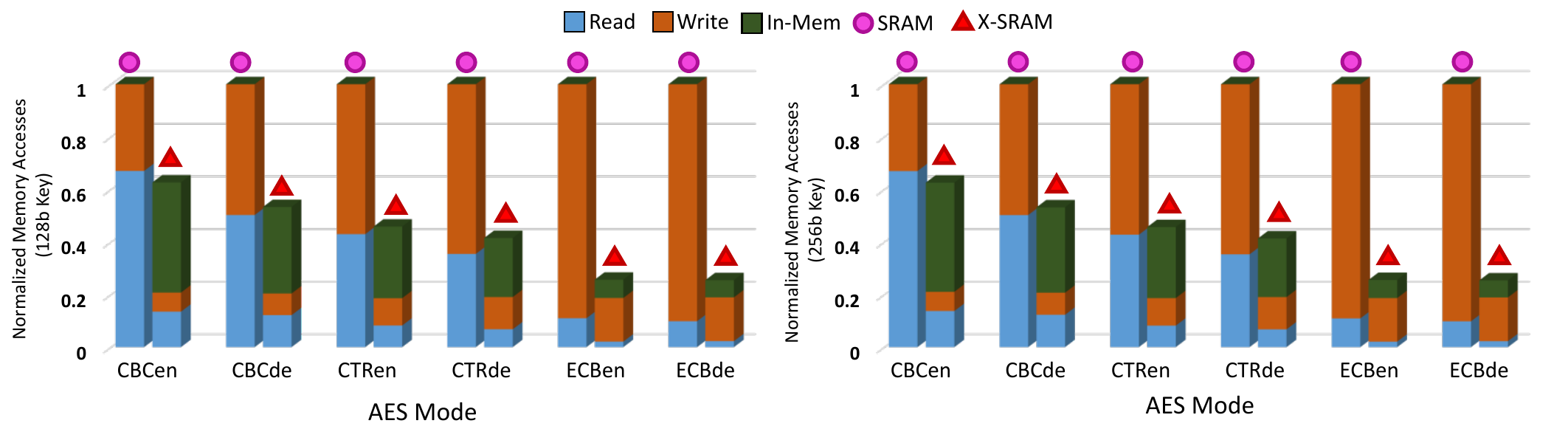}
\caption{Normalized number of memory accesses for various AES encryption and decryption modes and two different key-sizes, with and without using X-SRAM custom in-memory instructions. The total memory transactions are split into memory read instructions, memory write instructions and custom in-memory instructions.}
\label{aes_results}
\end{figure*}

In this section, we evaluate the system-level implications of using X-SRAMs instead of conventional SRAMs as the memory blocks in a typical von-Neumann based architecture taking advanced encryption algorithm (AES) as a case study. X-SRAMs enable extra functionalities within the memory block, as described in previous sections, through massively parallel vector Boolean operations. By utilizing such \textit{in-memory computations}, we expect reduction in energy expensive data movements over the bus between the processor and the memory blocks.

\subsection{Simulation Methodology}

A typical von-Neumann system implementation is shown in Fig \ref{systemimpl}(a). It consists of a processor, data-memory and an instruction-memory, connected by a system bus. For our simulations, we use Intel's programmable Nios-II processor \cite{nios2}, and extend the associated instruction set (ISA) to incorporate new custom-instructions enabled by our proposed X-SRAM (see Fig. \ref{systemimpl}(b)). The system bus follows the Avalon memory-mapped protocol, with enhanced architecture to enable passing three addresses at a time. Note that this is not a huge overhead since \textit{in-memory} instructions do not pass the data operands, and thus the data-channel along with the address-channel can be used to pass three memory addresses over the bus. This methodology is similar to the work presented in \cite{jain2}. A complete RTL model of the proposed X-SRAM was developed using the circuit parameters summarized in Table 1, incorporating the \textit{in-memory} computation capabilities. We perform cycle-accurate RTL simulations to run the benchmark AES application \cite{aes} on the architecture described above. AES encryption algorithm heavily relies on substitution-and-permutation operations that utilize several bit-wise Boolean operations such as XORs, which makes X-SRAM custom instructions suitable for this application. We identified pieces of code constituting 92\% of the entire runtime which can be mapped using the custom instructions RCS-XOR and RCS-Copy (shown in Fig. \ref{systemimpl}(b)), along with usual memory read-write instructions. The software was modified by replacing repetitive Boolean operations with our custom instruction macros.

\begin{figure}[t]
\centering
\includegraphics[width=0.5\textwidth]{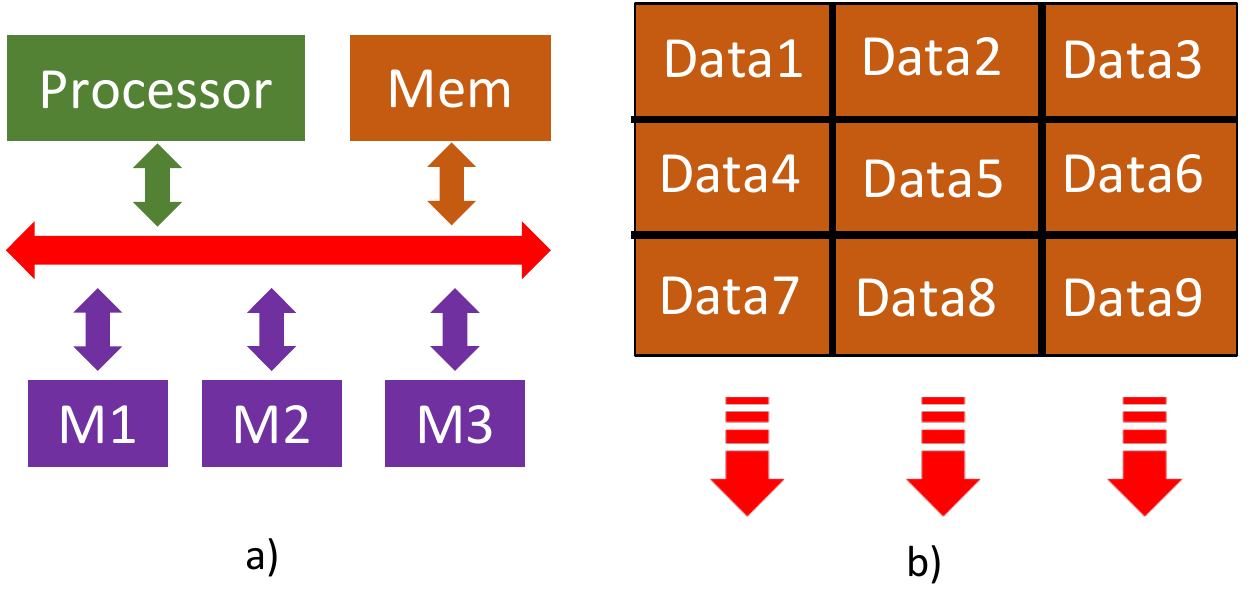}
\caption{(a) Realistic scenario for a typical system with multiple masters over a shared bus. An arbiter keeps track of the memory traffic and controls which master has access to the bus at a given point in time. (b) Data-parallelism in memory arrays. X-SRAM performs bit-wise operations throughout the row, where each row may store multiple data words. Thus multiple computations occur in parallel.}
\label{discussion}
\end{figure}

\subsection{Results and Discussion}

We evaluate three modes of AES encryption and decryption namely CBC, CTR and ECB \cite{cbcecb} for two different key sizes $-$ 128bits and 256bits. We plot the total number of memory accesses (memory read instructions, memory write instructions and custom in-memory instructions) required for each mode in Fig. \ref{aes_results}. The results are normalized to the corresponding memory accesses required in  a conventional SRAM memory block (no \textit{in-memory} custom instructions). The plots show that the memory accesses can be reduced by up-to 74.7\% and 74.6\% in ECB mode for 128b and 256b key respectively, by using X-SRAM in-memory instructions. The implications of these are threefold. 1) Since memory transactions are expensive, we directly save $\sim$75\% memory access energy consumption by reducing the number of accesses to the memory. The total energy consumption in the peripheral circuitry is also thereby reduced. 2) In a realistic scenario, shown in Fig. \ref{discussion}(a), multiple masters access the shared system bus, thereby causing large arbitration delays. Reducing the total number of memory accesses allows the system bus to cater to other masters, thereby reducing arbitration wait times over the shared bus and hence improving overall system performance. 3) The decrease in the data transfer volume between the processor and memory alleviates the problems associated with limited bus bandwidth while providing enough memory bandwidth for parallelism. Fig. \ref{discussion}(b) shows how data can be mapped to the X-SRAM to exploit data-parallelism. Since X-SRAM has capability to compute two physical rows at a time, Data1-4, Data2-5 and Data3-6 can be computed in parallel with a single in-memory instruction, thereby improving throughput.

\section{Conclusion}

Von-Neumann machines have fueled the computing era for the past few decades. However, the recent emphasis on data intensive applications like artificial intelligence, image recognition, cryptography \textit{etc.} requires novel computing paradigm in order to fulfill the energy and throughput requirements. `In-memory' computing has been proposed as a promising approach that could sustain the throughput and energy requirements for future computing platforms. In this paper, we have proposed multiple techniques to enable in memory computing in standard CMOS bit-cells $-$ the 8T cell and the 8$^+$T cell. We have shown that Boolean functions like NAND, NOR, IMP and XOR can be obtained by minimal changes in the peripherals circuits and the associated read-operation. Further, we have also proposed a \textit{`read-compute-store'} scheme by leveraging the decoupled read and write ports of the 8T and 8$^+$T cells, wherein the computed logic data can be directly stored in the desired row of the memory array. Our results are supported by rigorous Monte-Carlo simulations performed using predictive transistor models. Moreover, taking an example of AES encryption algorithm, we demonstrate that up-to 75\% memory transactions can be avoided, thereby allowing energy and performance improvements.

%\section*{Acknowledgements}
%The research was funded in part by C-BRIC, one of six centers in JUMP, a Semiconductor Research Corporation (SRC) program sponsored by DARPA,  the National Science Foundation, Intel Corporation and Vannevar Bush Faculty Fellowship.

\bibliographystyle{IEEEtran}
\bibliography{proj}

\begin{IEEEbiography}[{\includegraphics[width=1in,height=1.25in,clip,keepaspectratio]{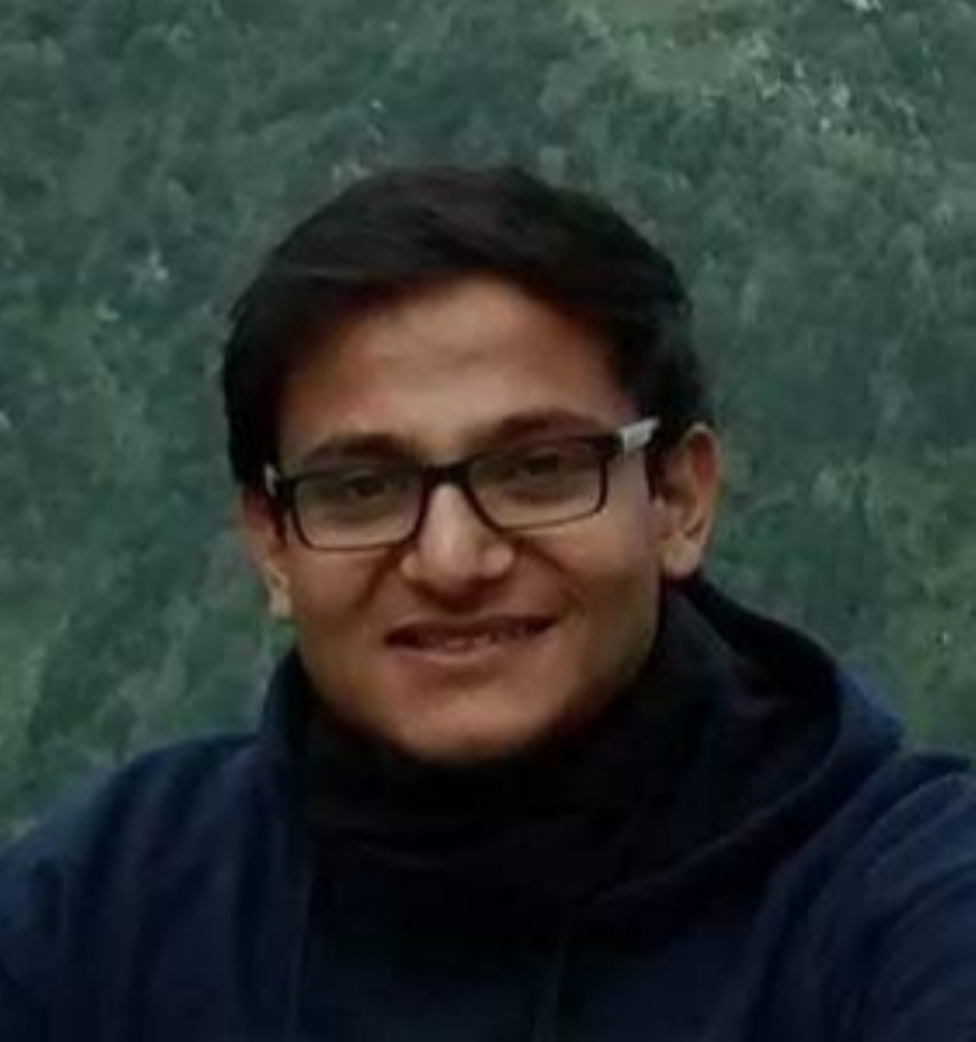}}]{Amogh Agrawal}
received his B.Tech degree in Electrical Engineering from Indian Institute of Technology (Ropar), India in 2016. He was a research intern at University of Ulm, Germany in 2015, under the DAAD (German Academic Exchange Service) fellowship. He joined the Nanoelectronics Research Lab in 2016 and is currently pursuing Ph.D. degree at Purdue University under the guidance of Prof. Kaushik Roy. His primary research interests include enabling in-memory computations for neuromorphic systems using CMOS and beyond-CMOS memories. He is also looking into modeling and simulation of emerging spintronic devices for applications in neuromorphic computing. He was the recipient of Director’s Gold Medal for his all-round performance, and Institute Silver Medal for his academic achievements at IIT Ropar. He is also the recipient of the Andrews Fellowship from Purdue University since 2016.
\end{IEEEbiography}

\begin{IEEEbiography}
[{\includegraphics[width=1in,height=1.25in,clip,keepaspectratio]{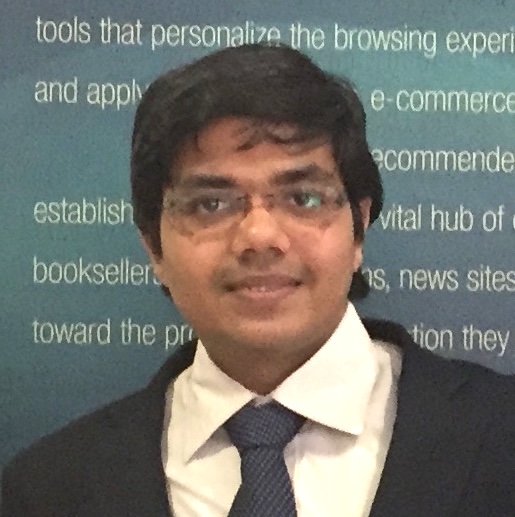}}]{Akhilesh Jaiswal} received the B.Tech degree from Shri Guru Gobind Singhji Institute of Engineering and Technology, Nanded, India, in 2011 and the M.S. degree in electrical engineering from the University of Minnesota, Minneapolis, MN, USA, in 2014. He joined Nano-electronics Research Lab at Purdue University in the Fall of 2014, where he is currently pursuing doctoral degree. He was an intern at Globalfoundries Lab, Malta, USA during summer 2017. His research interests include in-memory CMOS and beyond-CMOS computing, exploration and modeling of spin devices for on-chip memory/logic/nueromorphic applications.
\end{IEEEbiography}

\begin{IEEEbiography}
[{\includegraphics[width=1in,height=1.25in,clip,keepaspectratio]{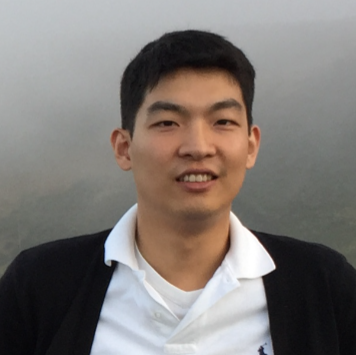}}]{Chankyu Lee} received B.S. in Electrical and Electronics Engineering from Sungkyunkwan University, Korea, in 2015. Currently, he is pursuing PhD degree in Electrical and Computer Engineering at Purdue University, West Lafayette, IN, USA. His primary research lies in the area of brain-inspired (neuromorphic) computing and event-driven deep learning, low power and high performance VLSI design for machine learning hardware.
\end{IEEEbiography}

\begin{IEEEbiography}[{\includegraphics[width=1in,height=1.25in,clip,keepaspectratio]{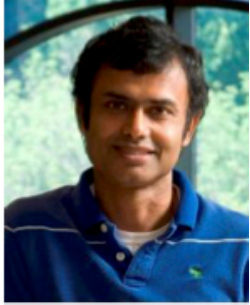}}]{Kaushik Roy}
received the BTech degree in electronics and electrical communications engineering from the Indian Institute of Technology, Kharagpur, India, and the PhD degree from the Department of Electrical and Computer Engineering, University of Illinois at Urbana-Champaign in 1990. He was with the Semiconductor Process and Design Center of Texas Instruments, Dallas, where he worked on FPGA architecture development and low-power circuit design. He joined the electrical and computer engineering faculty at Purdue University, West Lafayette, IN, in 1993, where he is currently Edward G. Tiedemann Jr. Distinguished Professor. His research interests include neuromorphic and cognitive computing, spintronics, device-circuit co-design for nano-scale Silicon and non-Silicon technologies, low-power electronics for portable computing and wireless communications, and new computing models enabled by emerging technologies. He has published more than 600 papers in refereed journals and conferences, holds 15 patents, graduated 70+ PhD students, and is coauthor of two books on Low Power CMOS VLSI Design (Wiley \& McGraw Hill). He received the US National Science Foundation Career Development Award in 1995, IBM faculty partnership award, ATT/Lucent Foundation award, 2005 SRC Technical Excellence Award, SRC Inventors Award, Purdue College of Engineering Research Excellence Award, Humboldt Research Award in 2010, 2010 IEEE Circuits and Systems Society Technical Achievement Award, Distinguished Alumnus Award from Indian Institute of Technology, Kharagpur, Fulbright Nehru Distinguished Chair, and Best Paper Awards at 1997 International Test Conference, IEEE 2000 International Symposium on Quality of IC Design, 2003 IEEE Latin American Test Workshop, 2003 IEEE Nano, 2004 IEEE International Conference on Computer Design, 2006 IEEE/ACM International Symposium on Low Power Electronics \& Design, and 2005 IEEE Circuits and System Society Outstanding Young Author Award (Chris Kim), 2006 IEEE Transactions on VLSI Systems Best Paper Award, 2012 ACM/IEEE International Symposium on Low Power Electronics and Design Best Paper Award, 2013 IEEE Transactions on VLSI Best Paper Award. He was a Purdue University Faculty scholar (1998-2003). He was a Research Visionary board member of Motorola Labs (2002) and held the M.K. Gandhi Distinguished Visiting faculty at Indian Institute of Technology (Bombay). He has been in the editorial board of IEEE Design and Test, IEEE Transactions on Circuits and Systems, IEEE Transactions on VLSI Systems, and IEEE Transactions on Electron Devices. He was the guest editor for Special Issue on Low-Power VLSI in the IEEE Design and Test (1994) and IEEE Transactions on VLSI Systems (June 2000), IEE Proceedings—Computers and Digital Techniques (July 2002), and IEEE Journal on Emerging and Selected Topics in Circuits and Systems (2011). He is a fellow of the IEEE.
\end{IEEEbiography}

\clearpage

\section*{Appendix}

\subsection{6-Transistor SRAM: bit-wise NOR/NAND/XOR Operation}
\label{sec:6t}

\begin{figure}[t]
\centering
\includegraphics[width=0.5\textwidth]{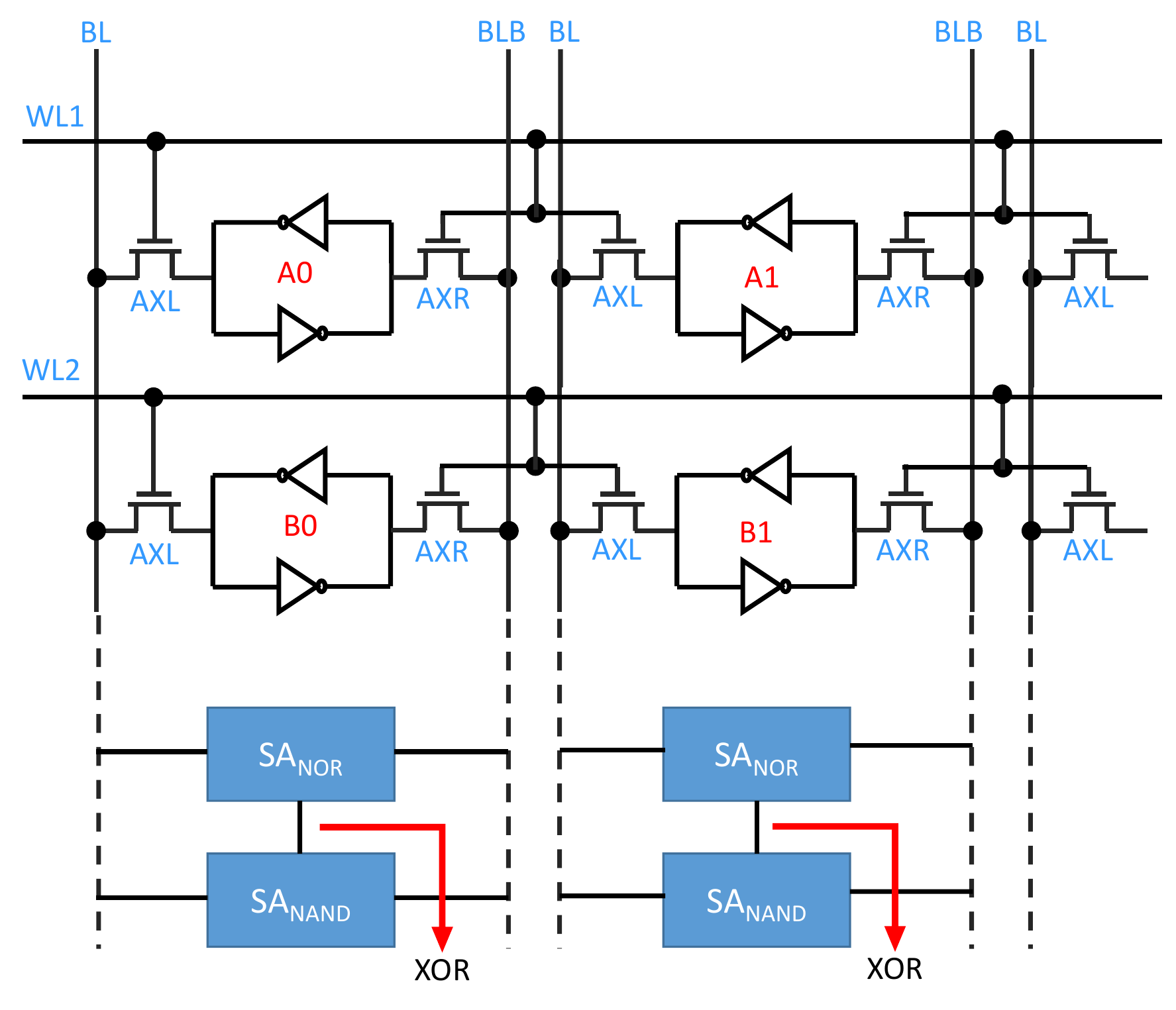}
\caption{Schematic of a 6T-SRAM array along with two asymmetric SAs in parallel for reading bitwise NAND/NOR/XOR operation.}
\label{fig:6tarr}
\end{figure}

\begin{figure}[t]
\centering
\includegraphics[width=0.5\textwidth]{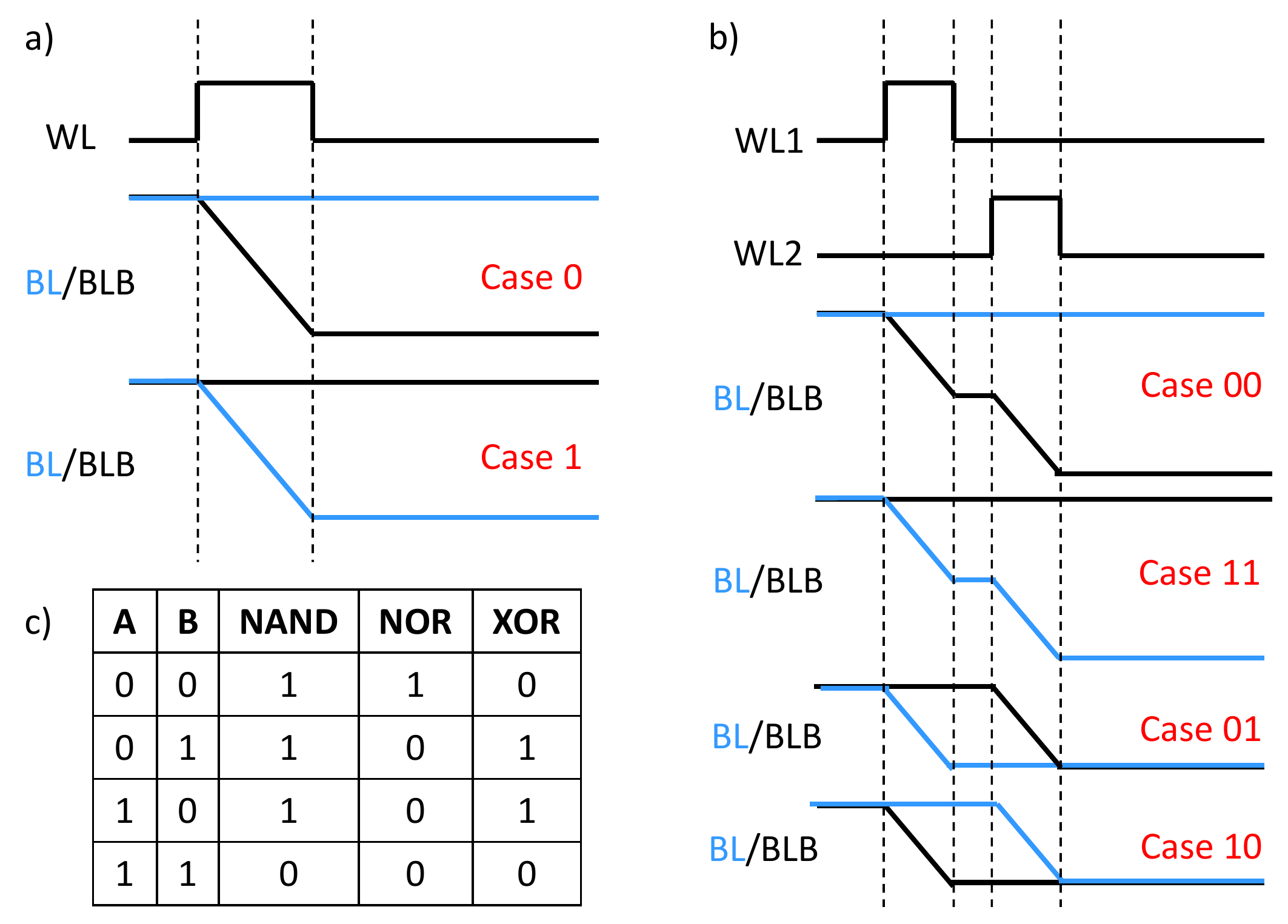}
\caption{a) Timing diagram for a typical memory read operation. The BL/BLB is pre-charged to Vdd, and the final voltage is shown for the two cases when the bit-cell stores `0' or `1'. b) Timing diagram for the proposed sequentially pulsed WL activation, and the resulting BL/BLB voltages for the four cases when bit-cells store `00,01,10,11'. c) Truth table for NAND/NOR/XOR operation.}
\label{fig:6t_timing}
\end{figure}

The most popular and widely used SRAM design is the standard 6T bit-cell, shown in Fig. \ref{fig:6tarr}. However, 6T bit-cells are inherently design constrained due to the shared read and write paths. Nevertheless, by proper design choices, 6T cells can still be used to perform in-memory computations although at reduced robustness due to the conflict between read and write operations in a standard 6T cell. The usual memory read operation in a 6T cell is performed by pre-charging the bit-lines (BL and BLB) to V$_{DD}$, and enabling the word-line corresponding to the row to be read out. Depending on whether the bit-cell stores `1' or `0', BL or BLB discharges, as illustrated in Fig. \ref{fig:6t_timing}(a). The difference in voltages on BL and BLB is sensed using a differential sense amplifier. 

Consider a typical memory array shown in Fig. \ref{fig:6tarr}, with two words `A' and `B' stored in rows 1 and 2, respectively. Simultaneously enabling WL1 and WL2 introduces read-disturbs due to possible short-circuit paths. Hence, we employ a sequentially pulsed WL technique as a workaround, similar to the proposal in \cite{shanbhag}. The address decoder sequentially turns WL1 and WL2 ON, corresponding to the rows storing `A' and `B', respectively, as illustrated in Fig. \ref{fig:6t_timing}(b).

The WL pulse duration is chosen such that with application of one WL pulse, BL/BLB drops to about $\sim$V$_{DD}$/2. If bits `A' and `B' both store `0' (`1'), BL (BLB) will finally discharge to 0V after the two consecutive pulses, whereas BLB (BL) remains at V$_{DD}$. On the other hand, for cases where `AB' = `10' and `01', the final voltages at BL and BLB would be the same ($\sim$V$_{DD}$/2), approximately. Thus, for the cases `01' and `10'  both BL and BLB would have a voltage $\sim$V$_{DD}$/2, while for `00' BL would be lower than BLB by $\sim$V$_{DD}$ and for the case of `10' BLB would be lower than BL by $\sim$V$_{DD}$.

\begin{figure*}[t]
\centering
\includegraphics[width=7in, height = 2in]{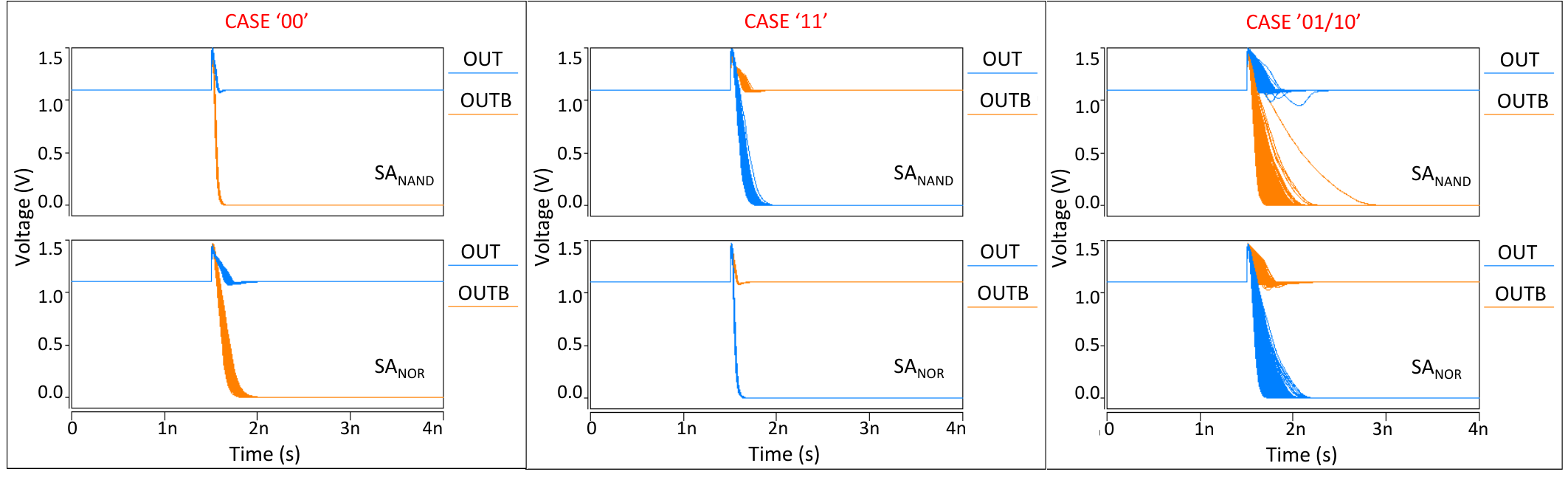}
\caption{Monte-Carlo simulations in SPICE of the SA outputs for all possible input cases $-$ `00,01,10,11', in presence of 30mV sigma variations in threshold voltage.}
\label{6t_mc}
\end{figure*}

The four cases are summarized in Fig. \ref{fig:6t_timing}(b). Using the two asymmetric SAs (in a similar fashion as proposed in Section \ref{8tdiff}), connected in parallel, we can obtain NAND/AND, NOR/OR and XOR bit-wise operations on `A' and `B'. The average energy consumption per-bit and the latency of in-memory operations in 6T-SRAM cells are 29.3fJ and 3ns, respectively. Note, although the sensing operation for in-memory computing with 6T cells seem similar to the 8$^+$T cell, there are certain key differences. Firstly, two word-lines cannot be activated simultaneously in 6T cells, therefore the WL pulses have to be properly timed and the pulse duration needs to be appropriately selected for achieving the desired functionality. Secondly, unlike the 6T cells the voltage swing on the read bit-lines for the 8$^+$T cells can have much larger swing without any concerns of possible read disturb failures, thereby relaxing the constraints on the sense amplifier.

A Monte-Carlo simulation with a 30mV sigma variations in the threshold voltage were performed to demonstrate the functionality and robustness of the proposal. Fig. \ref{6t_mc} shows the outputs of the asymmetric SAs, $SA_{NAND}$ and $SA_{NOR}$, for the four possible input cases - `00,01,10,11', in presence of variations. 

%Although 6T-SRAMs have been perfected over the years for reduced read/write errors, there are certain challenges faced by the proposed techniques for in-memory computations. Firstly, the read and write mechanisms impose conflicting constraints on the design of the 6T cell, thereby raising issues of read-disturbs due to coupled read-write paths. Secondly, the read disturb failure is accentuated by the fact that once the BL has discharged to V$_{DD}$/2, application of subsequent voltage pulse performs a \textit{pseudo-write} operation on the 6T cell, given the shared read and write paths. Thus, in the next two sections, we focus on 8T-SRAM cells, which enable decoupled read/write paths, thereby alleviating possible read-disturbs. Nevertheless, in the appendix, we show how the coupled read-write paths of the 6T cell can be exploited to implement a  simple `copy' operation, using boosted voltage techniques.

\subsection{6T SRAM: Copy Operation}

\begin{figure*}[t]
\centering
\includegraphics[width=0.8\textwidth]{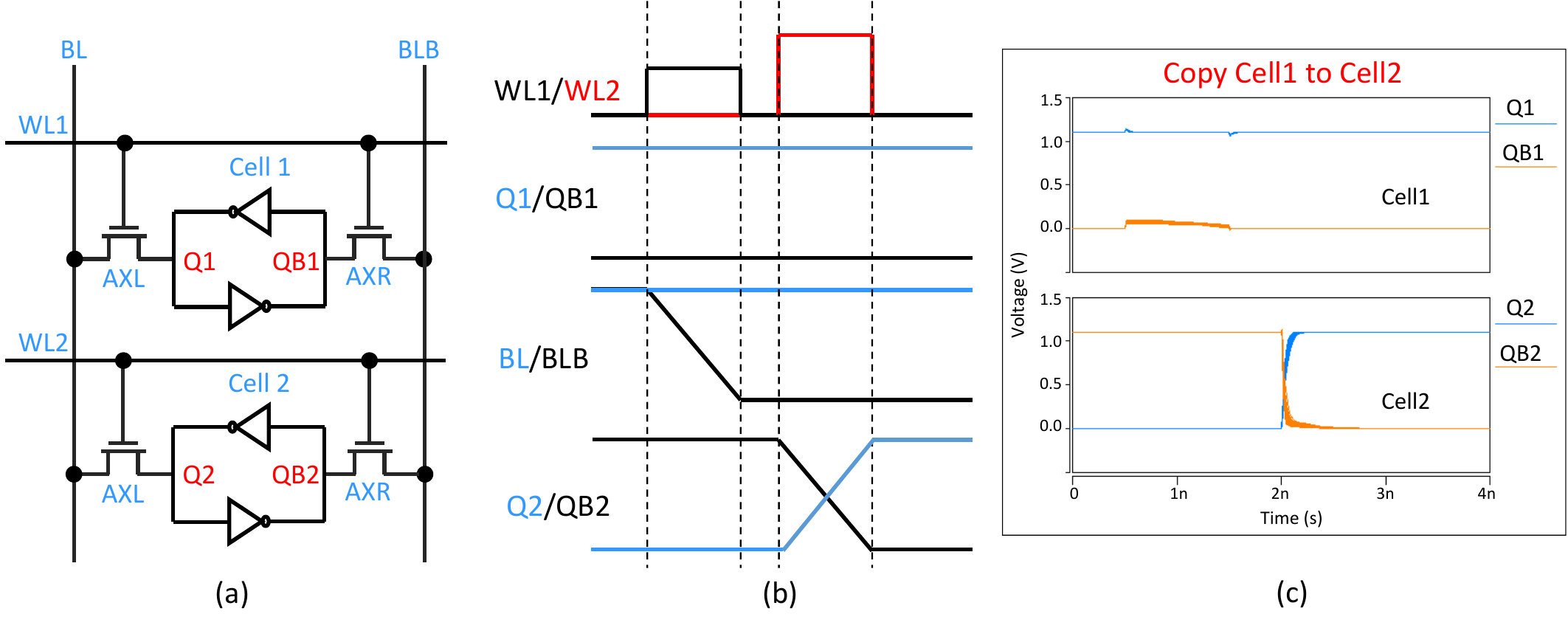}
\caption{a) Schematic of 6T-SRAM bit-cells $-$ Cell 1 and Cell 2. b) Timing diagram for performing a copy operation to copy data from Cell 1 to Cell 2. c) Monte-Carlo simulations in SPICE showing the final state of the Cell 2.}
\label{6t_appendix}
\end{figure*}

In this section, we describe a method of implementing `copy' functionality within the 6T bit-cell. To copy data from one memory location to another, a typical instruction sequence performed by the processor would be to do a memory read from the source location, followed by a memory write to the destination. Thus, two memory transactions are performed. We exploit the coupled read-write paths of the 6T cell to perform a data copy operation from one row to another, since the same set of bit-lines BL/BLB are used to read from and write into the cell.

Let us consider bit-cells 1 and 2, connected to WL1 and WL2, respectively (see Fig. \ref{6t_appendix}(a)). To copy data from cell 1 to cell 2, we perform two steps illustrated in Fig. \ref{6t_appendix}(b). Let us assume cell 1 stores a `1', while cell 2 stores `0'. The bit-lines BL/BLB are pre-charged to V$_{DD}$, as usual. In step 1, WL1 is enabled, thereby turning the access transistors of cell 1 ON. Since node Q1 is connected to V$_{DD}$ and QB1 is connected to 0V (cell 1 stores `1'), BL remains at V$_{DD}$, while BLB discharges to 0V. The pulse width is long enough for BLB to discharge fully to 0V. In step 2, WL2 is enabled, turning access transistors of cell 2 ON, with BL at V$_{DD}$ and BLB at 0V. Since cell 2 stores a `0', charge flows from BL to Q2, and from QB2 to BLB, thereby, flipping the state of cell 2, such that cell 2 now stores a `1'. If cell 2 initially stored a `1', nothing happens in step 2, and the state remains the same. Thus, we have implemented a data copy from cell 1 to cell 2, in a single memory transaction.

Note that step 1 is a usual memory read operation, while step 2 is similar to a memory write operation. However, step 2 is a weak write mechanism since the charge stored on BL/BLB is used to switch the cell state. This may cause write failures. Thus, a boosted voltage on WL2 is required to ensure correct data is written into cell 2. A Monte-Carlo simulation with sigma threshold voltage variations of 30mV in 45-nm PTM models was performed to test the proposal, as shown in Fig. \ref{6t_appendix}(c).

In order to implement a copy in 8T- and 8$^+$T Differential SRAMs, the proposed scheme would not work due to decoupled read-write paths. However, the RCS scheme proposed in Section \ref{sec:8t} can be used. The RWL of the source row and the WWL of the destination row are enabled, and the output of the sense amplifier is fed to the RCS block. This copies the data from the source row to the destination row in a single cycle operation.

\end{document}